\documentclass[aps,prl,reprint,superscriptaddress,amsmath,amssymb,floatfix]{revtex4-2}
\usepackage{graphicx}
\usepackage{dcolumn}
\usepackage{bm}
\usepackage{xcolor}

\begin{document}
\title{Three-dimensional flat bands and possible interlayer triplet pairing superconductivity in the alternating twisted NbSe$_2$ moir\'e bulk}
\author{Shuang Liu}
\affiliation{School of Physics and Electronics, Hunan University, Changsha 410082, China}
\affiliation{Department of Physics, Guangdong Technion - Israel Institute of Technology, 241 Daxue Road, Shantou, Guangdong 515063, China}
\author{Peng Chen}
\affiliation{Department of Physics, Guangdong Technion - Israel Institute of Technology, 241 Daxue Road, Shantou, Guangdong 515063, China}
\affiliation{Department of Physics, Technion – Israel Institute of Technology, 32000 Haifa, Israel}
\affiliation{Guangdong Provincial Key Laboratory of Materials and Technologies for Energy Conversion, Guangdong Technion – Israel Institute of Technology, Guangdong 515063, China}
\author{Shihao Zhang}
\email{zhangshh@hnu.edu.cn}
\affiliation{School of Physics and Electronics, Hunan University, Changsha 410082, China}
\begin{abstract}
Moiré superlattices hosting flat bands and correlated states have emerged as a focal topic in condensed matter research. Through first-principles calculations, we investigate three-dimensional flat bands in alternating twisted NbSe$_2$ moiré bulk structures. These structures exhibit enhanced interlayer interactions compared to twisted bilayer configurations. Our results demonstrate that moiré bulks undergo spontaneous large-scale structural relaxation, resulting in the formation of remarkably flat energy bands at twist angles $\leq$ 7.31°. The $k_z$-dependent dispersion of flat bands across different moiré bulks reveals their intrinsic three-dimensional character. The presence of out-of-plane mirror symmetry in these moiré bulk structures suggests possible interlayer triplet superconducting pairing mechanisms that differ from those in twisted bilayer systems. Our work paves the way for exploring potential three-dimensional flat bands in other moir\'e bulk systems.

\end{abstract}
\maketitle

\section{\label{sec:level1}INTRODUCTION}

Superconductivity in electronic systems with interacting electrons is a long-standing and captivating issue in condensed matter physics. Recent breakthroughs in twisted bilayer graphene (TBG) have underscored the critical role of engineered electronic band structures, particularly flat bands near the Fermi level, in facilitating unconventional superconductivity and correlated insulating states\cite{SC1,TBG1,TBG2}. This has propelled the field of twistronics, in which controlled rotational misalignment in van der Waals (vdW) structures creates moiré patterns that dramatically modify electronic structure into flat bands. Within these systems featuring flat bands, the electron-electron Coulomb interaction prevails over the kinetic energy\cite{TBG1,TBG2,flat1,flat2,flat3,flat4}, and thus provide a fertile platform for studying the interplay and competition of superconductivity and correlation\cite{FCI,QHE,torma2022superconductivity,lee2019theory,wu2018theory,marc-tbg-19, efetov-nature19, efetov-nature20,young-tbg-np20,li-tbg-science21,cao-tbg-nematic-science21,kang-tbg-prl19,Uchoa-ferroMott-prl,xie-tbg-2018, zaletel-tbg-2019,PhysRevLett.128.026403,PhysRevLett.128.247402,zaletel-tbg-hf-prx20,jpliu-tbghf-prb21,zhang-tbghf-arxiv20,hejazi-tbg-hf,kang-tbg-dmrg-prb20,kang-tbg-topomott,meng-tbg-arxiv20,Bernevig-tbg3-arxiv20,Lian-tbg4-arxiv20,regnault-tbg-ed,zaletel-dmrg-prb20,meng-tbg-qmc-cpl21,bultinck-tbg-strain-prl21}. 


In the twistronics, transition metal dichalcogenide (TMD) family has garnered significant attention\cite{xu2024maximally,mao2024transfer,zhang2021electronic,PhysRevLett.134.066601,PhysRevX.13.031037,fan2024orbital,tong2017topological,PhysRevLett.131.246501,PhysRevLett.133.066601,xu2025interplay,zhang2024polarization,PhysRevB.108.085117,PhysRevLett.128.026402,PhysRevLett.129.056804,PhysRevB.111.125122,PhysRevX.13.041026,PhysRevLett.122.086402,PhysRevB.109.L041106,cai2023signatures}. Notably, there has been considerable interest in the three-dimensional spiral structure of TMD, involving screw dislocations, which has been successfully grown through chemical vapor deposition (CVD) experiments\cite{zhang2014three,ci2022breaking,fan2017broken,chen2025structure}. Theoretical investigations have unveiled the presence of three-dimensional (3D) flat bands in the graphite moir\'e bulk\cite{lu2024magic}, which show the possible novel states in other 3D moir\'e structures.

Within the TMD family, NbSe$_2$ exhibits superconducting and charge density waves in few layers and bulk. The critical temperature of the superconducting rises with an increase in the number of layers\cite{NbSe2_SC}, suggesting that twisted NbSe$_2$ bulk may possess a higher critical temperature compared to twisted few layers. The varying numbers of layers and twist angles in twisted NbSe$_2$ lead to a diverse structural transition, influencing interlayer coupling and electronic characteristics. Those highlight twisted NbSe$_2$ bulk as a promising candidate for investigating 3D flat band structures and superconductivity.

In this work, using first-principles methods, we systematically studied the evolution of atom displacements, band structures and bandwidths in twisted angel ranging from 13.17° to 4.41°. The moir\'e bulk experiences spontaneous large-scale structural relaxation and exhibits extremely flat energy bands when twisted angle is not larger than 7.31°. Furthermore, the flat bands in various moiré structures exhibit a dependence on the $k_z$ direction, indicate of three-dimensional flat bands in the alternating twisted NbSe$_2$ moir\'e bulk. Especially, the bandwidth at 6.00° is narrower than that of the 1.05° TBG. Moreover, the out-of-plane mirror symmetry in the moir\'e bulk results in a distinct interlayer triplet pairing for superconductivity, distinguishing it from the twisted NbSe$_2$ bilayer system. 

\begin{figure*}
\includegraphics[width=0.8\linewidth]{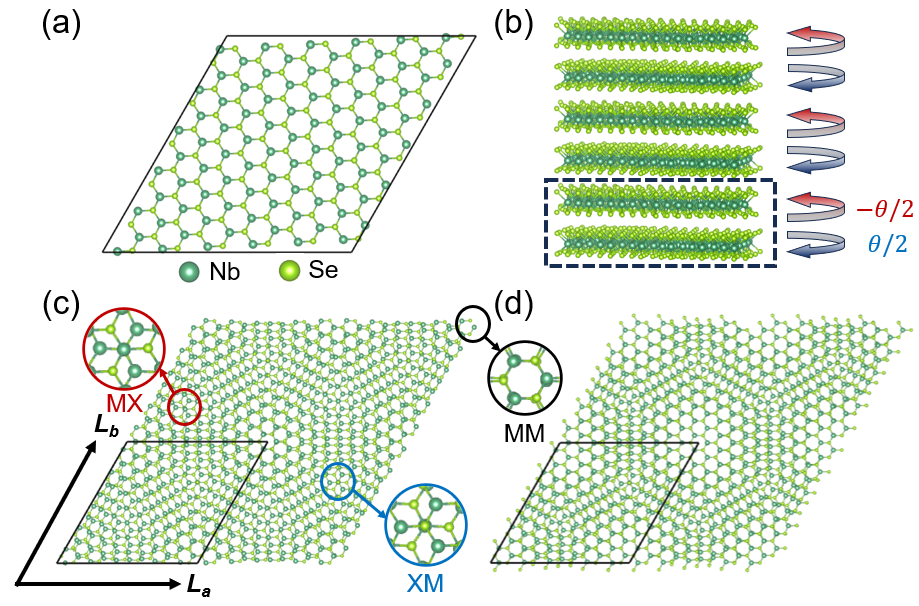}
\caption{(a) The top view of monolayer NbSe$_2$. (b) The side view of alternating twisted NbSe$_2$ moir\'e bulk. The unit cell of moir\'e bulk is remarked by black dashed lines. (c) The unrelaxed structure of alternating $6.01^{\circ}$-twisted NbSe$_2$ moir\'e bulk. (d) The relaxed structure of alternating $6.01^{\circ}$-twisted NbSe$_2$ moir\'e bulk. Compared to unrelaxed structure, obvious expansion of MM region exists in the relaxed structure.}
\end{figure*}

\section{\label{sec:level2}Calculations methods}
The structure optimizations of twisted NbSe$_2$ bulk were performed using the OpenMX with the Perdew–Burke–Ernzerhof (PBE) exchange–correlation function. We used the norm-conserving pseudopotentials and pseudo-atomic orbitals Se7.0-s3p2d2 and Nb7.0-s3p2d2\cite{OpenMX1,OpenMX2}. During the structure relaxation calculations, a k-points grid of 1$\times$1$\times$2 is adopted. The force criterion is set to $10^{-3}$ Hatree/Bohr. DFT-D3 method was employed to take into account vdW interactions during the structure optimizations\cite{OpenMX3}. 

The band structures of various twist angles were calculated by real-space electronic structure method implemented in the RESCU package\cite{RESCU}. In the self-consistent calculations, a k-points grid of 2$\times$2$\times$4 is adopted for twist angles ranging from 13.17° to 6.01° and 1$\times$1$\times$4 is adopted for 5.09° and 4.41°-twisted structures. The self-consistent procedure in the RESCU finish when the charge variation per valence electron is less than $10^{-6}$ e. \textcolor{black}{The reliability of RESCU is verified by comparing with VASP \cite{vasp_Kresse1993,vasp_Kresse1994,vasp_Kresse1996,vasp_Kresse1999} results within the bulk and large-angle twisted NbSe$_2$ as shown in Fig. S1.}

\section{\label{sec:level3}Results and discussions}
\subsection{Twisted bulk structures}
The crystal structure of NbSe$_2$ monolayer is illustrated in Fig.1 (a), with each layer lacking inversion symmetry. In this study, we consider the moiré bulk consistent of alternating twisted NbSe$_2$ bulk as depicted in Fig.1(b). In all calculations, the out-of-plane lattice parameter is fixed to relaxed parameter of aligned bilayer 13.4\AA{}. In these moiré bulks, only $C_{3z}$ rotational symmetry and $M_z$ mirror symmetry are preserved, while inversion symmetry and $M_y$ mirror symmetry are both broken.


\begin{table}[b]
\caption{\label{tab:table1}%
The relation between twisted angle, atoms $N_{atoms}$ and integers (m, n) in moiré unit cell. }
\begin{ruledtabular}
\begin{tabular}{cccc}
\textrm{n}&
\textrm{m}&
\textrm{angle}&
\textrm{$N_{atoms}$}\\
\colrule
2 & 3 & $ 13.17^{\circ}$ & 114\\
3 & 4 & $ 9.43^{\circ}$ & 222\\
4 & 5 & $7.34^{\circ}$ & 366\\
5 & 6 & $6.01^{\circ}$ & 546\\
6 & 7 & $5.09^{\circ}$ & 762\\
7 & 8 & $4.41^{\circ}$ & 1014\\
\end{tabular}
\end{ruledtabular}
\end{table}

\begin{figure*}
\includegraphics[width=0.8\linewidth]{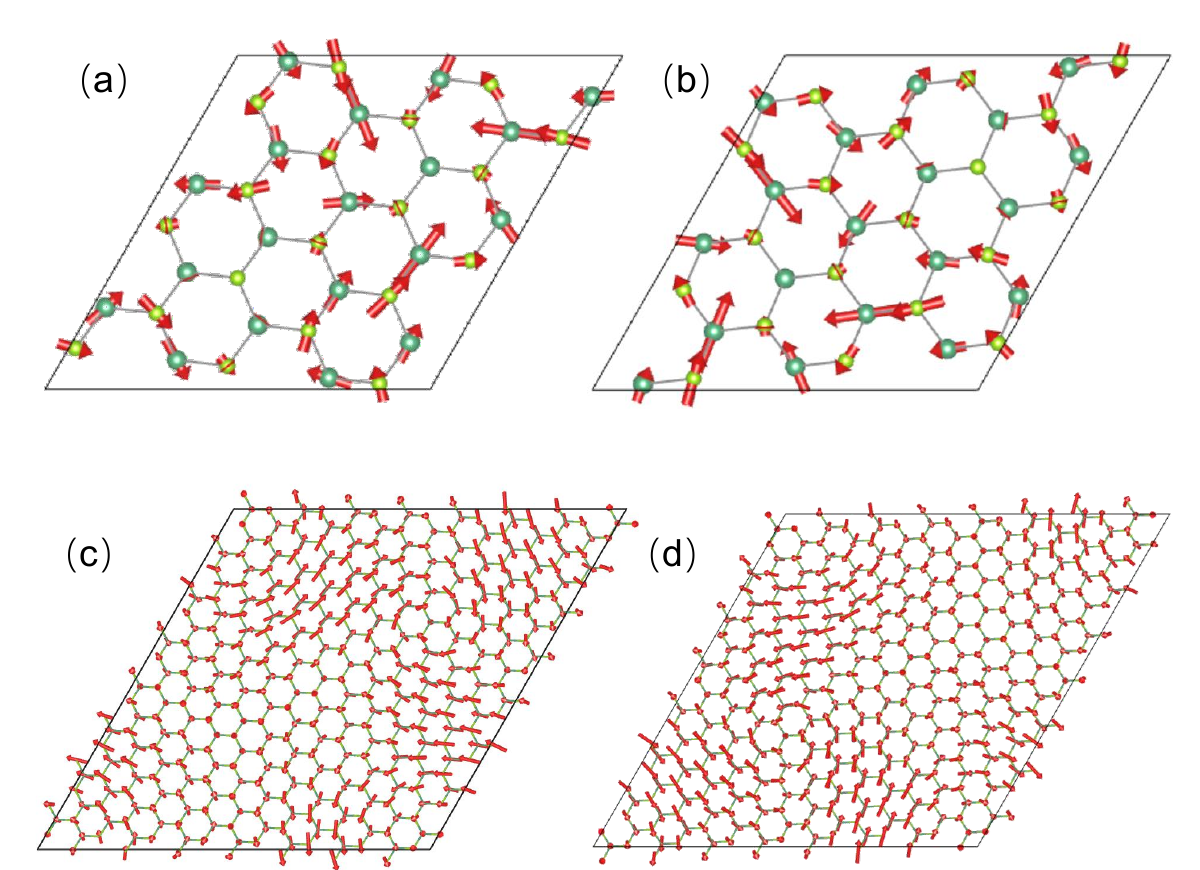}
\caption{The atomic displacements in the top layer (a) and bottom layer (b) in the unitcell of alternating $13.17^{\circ}$-twisted NbSe$_2$ moir\'e bulk are shown, the length of red vectors on each atom refers to atomic displacement amplitude \textcolor{black}{with scaling factor of 25}. Here atomic displacements in the top layer (c) and bottom layer (d) in the unitcell of alternating $4.41^{\circ}$-twisted NbSe$_2$ moir\'e bulk are also present \textcolor{black}{with scaling factor of 3}. \textcolor{black}{One can see Fig. S2 for more clear displacement distributions.} }
\end{figure*}

The lattice vectors of moiré unit cells can be descripted with a pair of integers ($m, n$)\cite{twisted1,twisted2}. One of the moiré lattice vectors can be obtained by $L_a = ml_a +nl_b$, where $l_a$ and $l_b$ are the lattice vectors corresponding to constituent monolayers. The commensurate twisted angle $\theta$ is related to ($m, n$),
\begin{equation}
    \cos \theta = \frac{1}{2}\frac{m^2+n^2+4mn}{m^2+n^2+mn}.
\end{equation}
The other moiré lattice vector $L_b$ is obtained by rotating $L_a$ by 60°. In our cases, we considered the twisted angle $\theta$ within the range of 4.41° to 13.17°. The relationship between twisted angle, number of atoms  and integers (m, n) in the moiré unit cell are listed in the Table.I. Fig. 1 (c) shows the unrelaxed 6.01°-twisted moiré pattern of NbSe$_2$ bulk from top view. There are three high-symmetry positions within the moiré unitcell, including MM, MX and XM staking patterns. The center of MM region is located at the original point, and the center of MX and XM region is $(L_a+L_b)/3$ and $2(L_a+L_b)/3$, respectively. Our relaxed structure is present in the Fig. 1(d), and we can note that MM region becomes remarkably expanded compared to unrelaxed cases, while MX and XM regions reduce to domain walls in the relaxed structure.

In the relaxed structure, the atomic displacement of $\theta$=13.17° and $\theta$=4.41° are shown in Fig. 2. In the large-angle ($\theta$=13.17°) twisted structure, only a few atoms experience remarkable displacements. In the top layer, clockwise displacements occur around the MX position. In the bottom layer, anti-clockwise displacements occur around the XM site. However, in the small-angle twisted structure with $\theta$=4.41°, distinctive vortex-like patterns of atomic displacements emerge, predominantly concentrated around the MM site, leading to the formation of a larger MM region within the moiré structure. \textcolor{black}{Each layer experiences interlayer interactions from both adjacent layers, which collectively suppress significant out-of-plane corrugation in the moiré bulk structure.}
The results of atomic displacement are consistent with Ref \cite{vortexCDW}, and reveal that remarkable vortex-like spontaneous displacement in the moiré bulk structure. \textcolor{black}{The large-angle $\theta \leq $7.34°) twisted bulk NbSe$_2$ system exhibits remarkable atomic relaxation, which is only present in the small-angle ($\theta \leq $3.14°) twisted bilayer NbSe$_2$\cite{vortexCDW}. This is the key viewpoint presented in our work, which emphasizes the important role of the twisted bulk in tuning atomic relaxation through strong interlayer interaction.}

\subsection{Electronic properties}
To study the three-dimensional electronic structure of twisted NbSe$_2$ bulk, we calculated the band structures on different $k_x-k_y$ planes across the Brillouin zone ($k_z$= 0, $\pi$/3, $\pi$/2). As shown in Fig. 3, the energy bands near Fermi level are all $k_z$-dependent in the alternating twisted NbSe$_2$ moir\'e bulk.

\begin{figure*}
\includegraphics[width=1.0\linewidth]{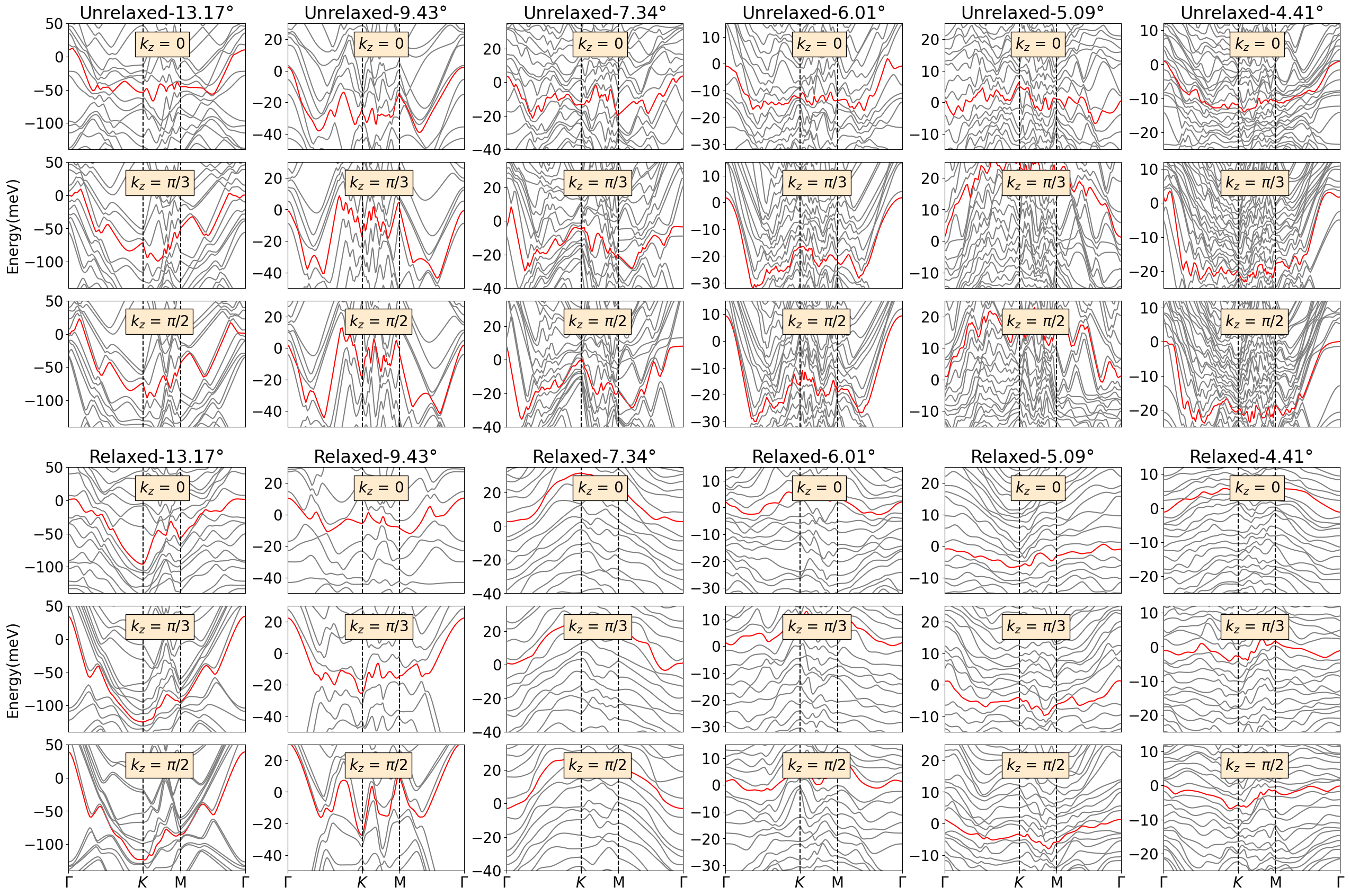}
\caption{The energy bands of alternating twisted NbSe$_2$ moir\'e bulk with unrelaxed and relaxed structures. Here we highlighted the energy bands whose energy is closest to Fermi level.}
\end{figure*}

In the unrelaxed structure of alternating twisted NbSe$_2$ moir\'e bulk with $\theta=13.17^{\circ}$, $9.43^{\circ}$, $7.32^{\circ}$ and $6.01^{\circ}$, there exist hole pockets located at $\Gamma$ point. However, in the relaxed structure, hole pockets are only maintained in the twisted NbSe$_2$ moir\'e bulk with $\theta=13.17^{\circ}$ and $9.43^{\circ}$ twist angles. This difference arises because energy bands become more dispersionless in the relaxed structure with smaller twist angle. Especially, large bandgaps at $\Gamma$-point disappear in the relaxed structure, and  more electronic states occur near the Fermi level in the alternating twisted NbSe$_2$ moir\'e bulk with twist angle $\theta \leq 7.34^{\circ}$. These results reveal that structural spontaneous relaxations in the alternating twisted NbSe$_2$ moir\'e bulk play an important role in their electronic structures. \textcolor{black}{The expansion of MM stacking regions and the contraction of MX/XM domains into domain walls confine the more localized electronic states in real space, which further suppresses kinetic energy and flattens the electronic structure.}

\begin{figure}
\includegraphics[width=1.0\linewidth]{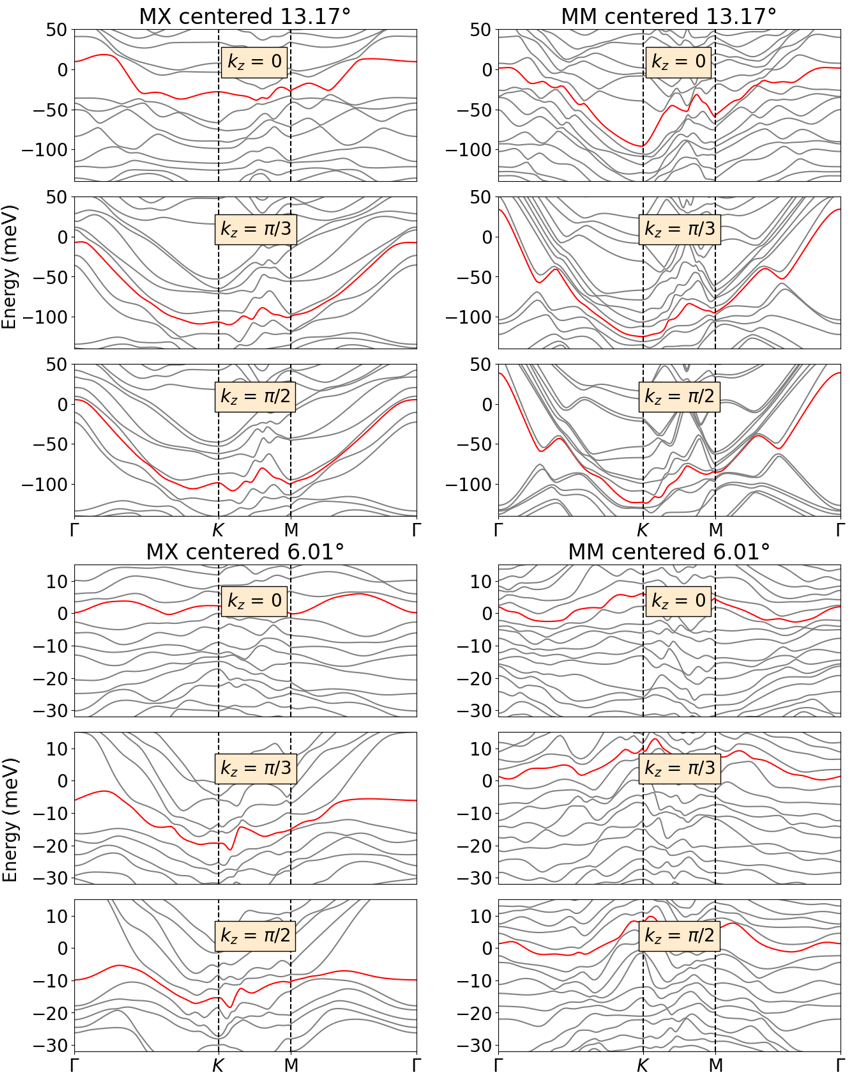}
\caption{Band structures in twisted 13.17° and 6.01° $NbSe_2$ with a rotation center of MX and MM (result of main text) respectively.}
\end{figure}

We note that lattice relaxation in the twisted bilayer graphene leads to gap opening at ±4 filling and particle-hole asymmetry\cite{TBG_gap_Guinea2019,TBG_gap_Miao2023,TBG_gap_Uchida2014,TBG_gap_Xie2023}. However, \textcolor{black}{it is observed that there is a dense set of mini-bands near the Fermi level, which stems from the conducting bands within the primitive unit-cell of the NbSe$_2$ monolayer.} The atomic relaxation fails to form similarly isolated flat bands in the twisted NbSe$_2$ bulk, which may be attributed to strong intralayer hopping and large intrinsic bandwidth of NbSe$_2$ compared to interlayer interaction.

\textcolor{black}{In the NbSe$_2$ monolayer, the energy bands in the $\Gamma$ valley originate from $d_{z^2}$ orbitals, but in the K valley, energy bands are mainly contributed by $d_{xy}+d_{x^2-y^2}$ orbitals\cite{NbSe2_orbit_Wickramaratne2020}. Orbital projection analysis for the $21.79^{\circ}$ twisted bulk (Fig. S4) confirms that the flat bands proximate to the Fermi energy exhibit K-valley character, dominated by the $d_{xy}+d_{x^2-y^2}$ orbitals.}

\begin{figure}
\includegraphics[width=1.0\linewidth]{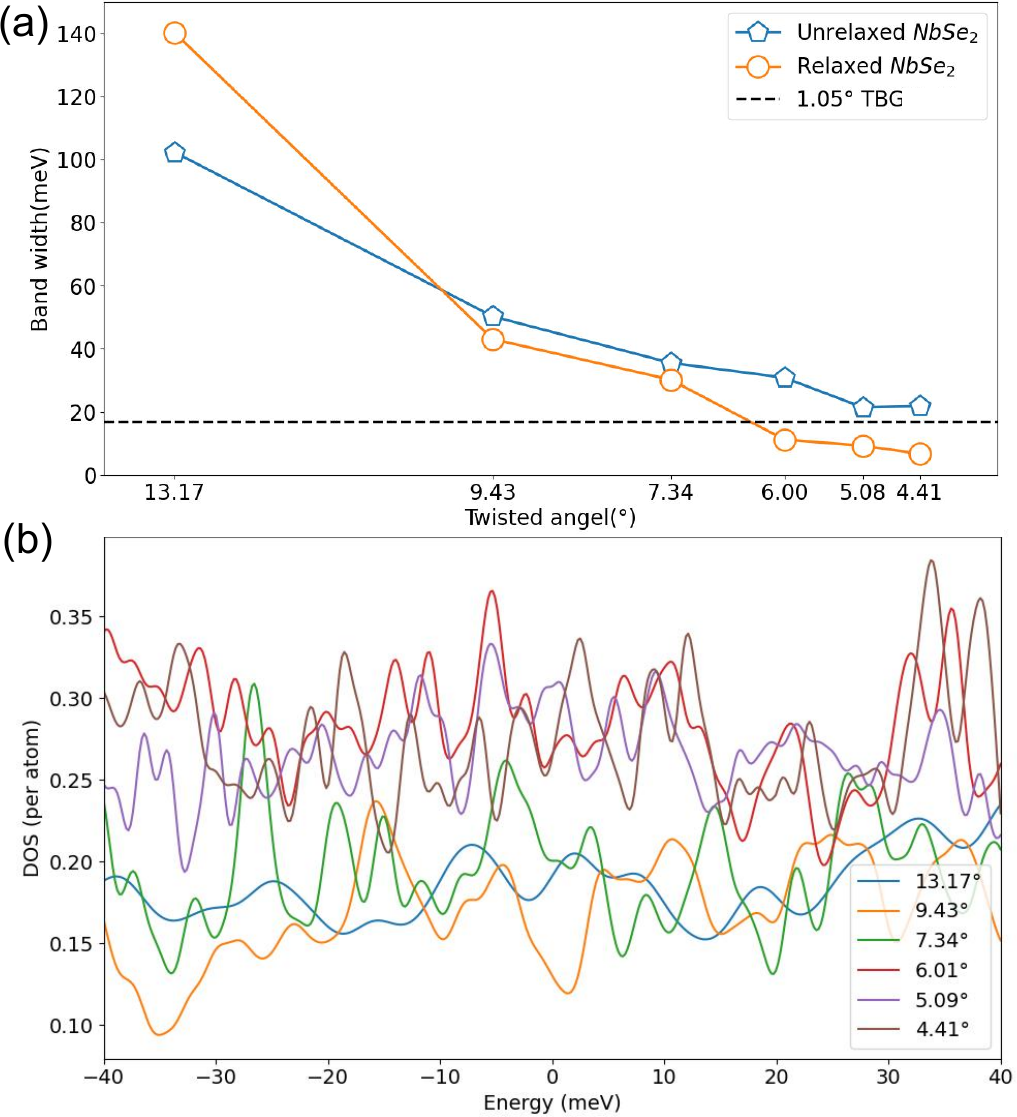}
\caption{(a) The average bandwidths of energy bands highlighted in the Fig.3 in different alternating twisted NbSe$_2$ moir\'e bulk. \textcolor{black}{(b) The calculated density of states per atom of different twisted NbSe$_2$ bulk.}}
\end{figure}

\textcolor{black}{It is noted that there are two different configurations in the twisted transition-metal dichalcogenides (TMD), which centers on the MM region and MX region\cite{stack_MX_Enaldiev2020}. In the MX-stacking twisted TMD, one layer is obtained by rotating by $180^{\circ}$ around the hollow compared to MM-stacking twisted TMD. Previous work\cite{stack_MX_Enaldiev2020} showed that the electronic structures behave very different in these twisted structure. Thus, we also calculated the electronic structure of $13.17^{\circ}$ and $6.01^{\circ}$ with AP centered twisted structure as shown in Fig. 4. Although the overall band shapes differ significantly between the MX- and MM-centered twisting structures, the bandwidths of the bands near the Fermi energy remain similar: 93.3 meV (MX) vs. 102.3 meV (MM) for $13.17^{\circ}$, and 12.6 meV (MX) vs. 11.1 meV (MM) for $6.01^{\circ}$. These results indicate that the bandwidths of the bands near the Fermi energy is mainly insensitive to the twisting center.}

To trace the evolution of the bandwidth with decreasing twist angle, we focus on the bands closest to the Fermi level. The average bandwidth of these bands at different $k_z$ is used as an indicator of the system's overall bandwidth. As shown in Fig.5, the bandwidths of alternating $7.34^{\circ}$-twisted NbSe$_2$ moir\'e bulk decrease to 30\,meV. Subsequently, in the alternating $6.01^{\circ}$-twisted NbSe$_2$ moir\'e bulk, the bandwidths are further reduced to 10\,meV, which are close to typical bandwidths of magic-angle twisted bilayer graphene as shown in the Fig.\,5. Thus, three-dimensional flat bands emerge in the alternating twisted NbSe$_2$ moir\'e bulk with twisted angle $\theta \leq 6.01^{\circ}$, which can experience significant  electron-electron interaction. \textcolor{black}{As shown in Fig.\,5, the atomic relaxation in the twisted moiré bulk magnifies the density of states (DOS) near the Fermi level. The small bandwidth of mini-bands induced by atomic relaxation results in a series of sharp DOS peaks in the 6.01°, 5.09,° and 4.41° twisted NbSe$_2$ bulk.}
In the alternating twisted NbSe$_2$ moir\'e bulk, each layer possesses interlayer interaction from two neighboring layers. But in the twisted bilayer NbSe$_2$ or at the surface of moir\'e bulk, these interlayer interactions weaken, leading to an enhancement in the bandwidth of moir\'e bands. Thus, the surface state behaves different from bulk states in the alternating twisted NbSe$_2$ moir\'e bulk.

\subsection{Electron Coulomb interaction}
The electron Coulomb interaction can be written as $H=V(q) c_{k+q}^{\dagger} c_{k^{\prime}-q}^{\dagger} c_{k^{\prime}} c_k$, where $k, k^{\prime}, q$ are wave vectors in the Brillouin zone. In the 2D cases, $V(q)=e^2 / (2 \Omega_M \epsilon \epsilon_0 \sqrt{q^2+\kappa^2})$ denotes the screened Coulomb interaction, where $\Omega_M$ is the area of moiré lattice, $\kappa$ is the inverse screening length, and $\epsilon$ denotes the background dielectric constant. But in the 3D cases, $V(q)=e^2 / [d \Omega_M \epsilon \epsilon_0\left(q^2+\kappa^2\right)]$ where $d$ refers to out-of-plane lattice constant. In the experiments, $\kappa=0.002 \sim 0.02 \AA^{-1}$. We can note that at the small $q$ (meets $d \sqrt{q^2+\kappa^2}<$
2), electron interaction in the 3D case is stronger than that in the 2D twisted system. In the 6.01$^\circ$ twisted NbSe$_2$ moiré bulk, when $q \sim 0.1 G=0.022 \AA^{-1}$ ( $G$ is in-plane reciprocal lattice vector length) and $\kappa=0.01 \AA^{-1}$, $\frac{2}{d\sqrt{q^2+\kappa ^2}} = 6.1762$ reveals that the 3D electron Coulomb interaction is about 6 times as large as that of 2D electron interaction. Especially, 3D electron Coulomb interaction delays rapidly with $\sim 1 / q^2$ power law, and we note that $V(q=0.1 G) / V(q=0.316 G)=10$ indicates of short-range electron interaction. Thus, the dominant electron Coulomb interaction in the 3D moiré system originates from small-$q$ interaction term. \textcolor{black}{In the quasi-two-dimensional twisted bilayer system, the two-dimensional electron Coulomb interaction decays gradually following the 1/$q$ power law, and inter-band scattering cannot be disregarded, which inhibits the twisted bilayer system from attaining correlated states. Nevertheless, in the twisted moiré bulk, the three-dimensional electron Coulomb interaction decays rapidly in accordance with the 1/$q^2$ power law. Despite the presence of a multiband electronic structure near the Fermi level, the dominant electron interactions are confined to small-q scattering, the region of which encompasses only one or two energy bands. Consequently, the characteristic of the three-dimensional electron Coulomb interaction eliminates the constraint regarding isolated flat bands in the two-dimensional twisted moiré system.}

In the case of alternating twisted trilayer graphene, flat bands coexist with highly dispersive Dirac cones\cite{Kim2023,Shen2023}. These bands overlap near the Fermi level. Despite this, experiments have observed correlated states within this moiré system. Theoretical work\cite{PhysRevX.12.021018} has also investigated these states, referring to them as intervalley coherent states. Therefore, the presence of correlated states in the alternating twisted trilayer graphene demonstrates that overlapping bands do not suppress the correlated state within the flat bands. Thus, multi-band scattering or excitation may not suppress the correlated states in our alternating twisted bulk.

\textcolor{black}{It is noted that the supercell is enlarged with the decrease of twist angle. At the same time, the effect of band folding is evident to reduce the bandwidths. We define the characteristic energy on the moir\'e length as $U_M=e^2 /\left(4 \pi \epsilon \epsilon_0 L_S\right)$\cite{PhysRevLett.128.026403}. The comparison between characteristic interaction energy and bandwidth can give out proper scale analysis. In the magic-angle twisted bilayer graphene, the characteristic energy is about 22 meV , which is larger than the bandwidth of flat bands. As for $6.01^{\circ}$ twisted NbSe$_2$ moir\'e bulk, the reciprocal lattice length is 4 times as large as that of magic-angle twisted bilayer graphene. Thus, the characteristic energy is about 88 meV , which is much larger than the bandwidths of energy bands. These scale analysis show that 6.01 twisted NbSe$_2$ moir\'e bulk resides in the strong interaction regime favorable for correlated electron physics.}

\subsection{Interlayer triplet pairing superconductivity}
 		

NbSe$_2$ always exhibits superconductivity in few layers and bulk. Here we discuss the possible superconductivity in the alternating twisted NbSe$_2$ moir\'e bulk. In the superconducting materials, the electrodynamic characteristics are expressed by
\begin{equation}
    \bm{j}=-D_s\bm{A}.
\end{equation}
Here $\bm{j}$ is current density, $\bm{A}$ denotes vector potential and $D_s$ refers to superfluid weight. The superfluid weight is crucial criterion of superconductivity. It can be divided into two sections: conventional contribution and geometric contribution\cite{torma2022superconductivity}. The conventional contribution is drived from single-band model, but geometric contribution originates from multi-band framework\cite{torma2022superconductivity}. In the flat-band system, geometric superfluid weight predominates over conventional superfluid weight. Thus, in the alternating twisted NbSe$_2$ bulk with smaller twist angles, geometric superfluid weight plays a more significant role. 
\textcolor{black}{While the increased effective mass suggests enhanced band flatness, a key indicator for superconductivity, it is crucial to note that the geometric superfluid weight depends not only on the effective mass but also sensitively on the geometric properties of the electronic wavefunctions\cite{Geo_flat_band_Han2024}. The influence of wavefunction geometry on superconductivity within the alternating twisted NbSe$_2$ bulk requires further dedicated investigation in future work.}
 
\textcolor{black}{Now we discuss possible supercoducting pairings in the alternating twisted NbSe$_2$ bulk. 
The continuum Hamiltonian for the superconducting part can be expressed as
$$
H_{s c}(\bm{r})=\int d \bm{r} \sum_{\xi l l^{\prime} s s^{\prime}} \psi_{\xi l s}^{\dagger}(\bm{r}) \Delta_{l l^{\prime}, s s^{\prime}}(\bm{r}) \psi_{-\xi l^{\prime} s^{\prime}}^{\dagger}(\bm{r})+h . c .
$$
Here $\xi$ refers to valley index and $l$ or $l'$ represents the layer in the unitcell of moiré bulk. Following the previous work\cite{SC_hetero_Gani2019,PhysRevLett.131.016001}, we only consider momentum-direction-independent interaction pairings in this work, for simplicity. The electronic band structure only indicates of single-particle wavefunction, so we need to determine the certain form of interaction pairing by symmetry constraint.}

To begin, we analyze the possible superconducting pairings in the twisted bilayer NbSe$_2$. In the twisted bilayer NbSe$_2$, the system exhibits $\mathcal{T}\times D_3$ symmetry\cite{PhysRevLett.131.016001}. Thus, twisted bilayer NbSe$_2$ has time-reversal symmetry $\mathcal{T}=is_y\mathcal{K}$, three-fold rotational symmetry $C_{3z}=e^{i2\pi s_z/3}$ and two-fold rotational symmetry $C_{2y}=-i\tau_xs_y$. Here $\tau$ and $s$ are Pauli matrices defined in layer and spin space, respectively. There are only six matrices of pairing wavefunction: $s_y$, $\tau_xs_y$, $\tau_zs_y$, $\tau_y$, $\tau_ys_x$, $\tau_ys_z$ can couple to the s-wave pairings. Among these possible pairings, only $ \{ \tau_y,\tau_ys_z\} $ pairing belongs to two-dimensional irreducible representation, and other pairings belong to one-dimensional irreducible representation. The $(\tau_0,\tau_x)s_y$, $\tau_zs_y$, and $\tau_ys_x$ describe the singlet pairings, and $ \{ \tau_y,\tau_ys_z\} $ refers to triplet pairing. However, in the alternating twisted NbSe$_2$ bulk, extra mirror symmetry $M_z=is_z$ rule out all pairings of one-dimensional irreducible representation. Thus, only $ \{ \tau_y,\tau_ys_z\} $ pairing is possible in the alternating twisted NbSe$_2$ bulk, whose explicit pairing form is

\begin{figure*}
\includegraphics[width=0.8\linewidth]{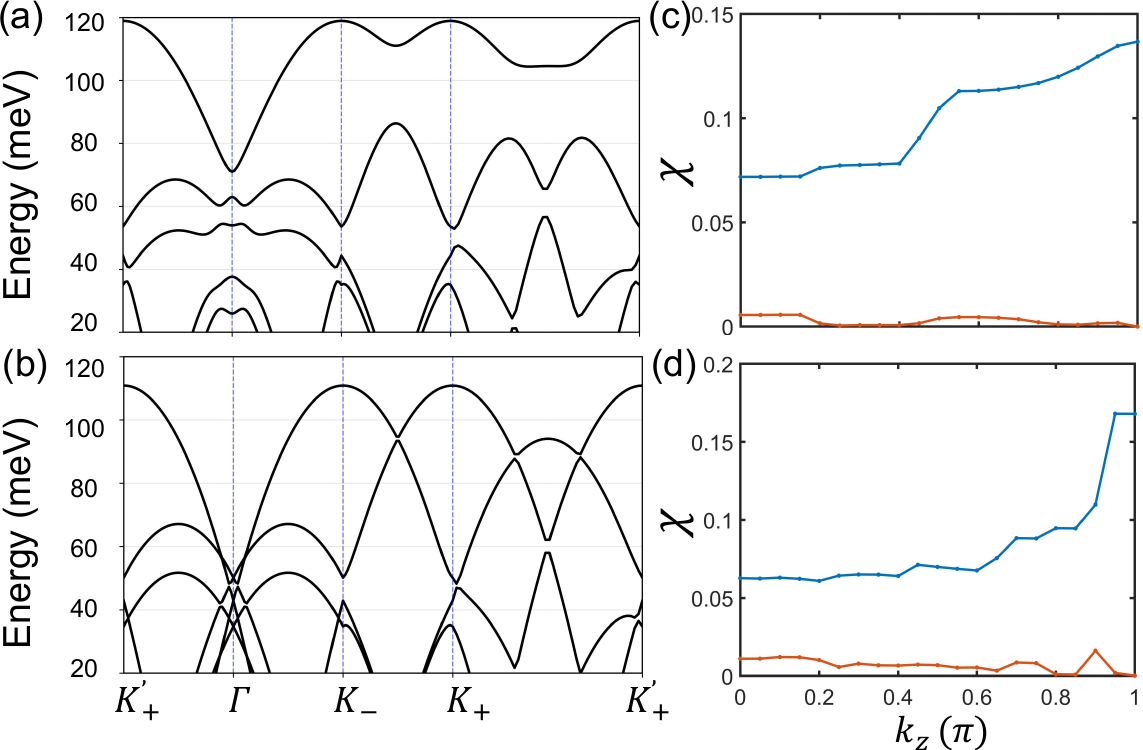}
\caption{\textcolor{black}{The energy bands of effective model under 5$^\circ$ twisted angle in the (a) $k_z=0$ and (b) $k_z=\pi$ plane. The calculated $k_z$-resolved pairing susceptibility at the (c) $\mu=50$\,meV and (d) $\mu=80$\,meV chemical potential. The blue lines refer to the $k_z$-resolved pairing susceptibility of $\gamma _{y0}=\tau_y$ channel. And the red lines represent the $k_z$-resolved pairing susceptibility of mixture between $\gamma _{y0}=\tau_y$ and $\gamma _{yz} = \tau_ys_z$ channel.}}
\end{figure*}

\begin{widetext}
\begin{equation}
    \Psi_{BCS}=\{\psi_{-\xi-\bm{k}\,,\,l\uparrow}\,\psi_{\xi+\bm{k}\,,\,l'\uparrow}+\psi_{-\xi-\bm{k}\,,\,l\downarrow}\,\psi_{\xi+\bm{k}\,,\,l'\downarrow}, \,\,\,\, \psi_{-\xi-\bm{k}\,,\,l\uparrow}\,\psi_{\xi+\bm{k}\,,\,l'\uparrow}-\psi_{-\xi-\bm{k}\,,\, l\downarrow}\,\psi_{\xi+\bm{k}\,,\, l'\downarrow}\} \,\,\,(l\neq l').
\end{equation}
\end{widetext}
Thus, only interlayer triplet pairing can exist in the alternating twisted NbSe$_2$ bulk, which is different from twisted bilayer system.

\textcolor{black}{Now we turn to discuss the microscopic model of the superconductivity in the twisted moiré bulk. We begin with the NbSe$_2$ monolayer, which has conducting bands across the Fermi level. Its electronic structure is different from those of the MoS$_2$ and MoTe$_2$ insulating monolayers. The continuum model of twisted MoS$_2$ and MoTe$_2$ can be accurately described by the electronic states near the K valley, but the accurate continuum model of twisted NbSe$_2$ needs to be described by the full Hamiltonian in the entire Brillouin zone. To discuss the microscopic model of the superconductivity in the twisted moiré bulk, we can start from the minimal model. Our calculations show that the electronic states in the twisted bulk NbSe$_2$ near the Fermi level originate from the electronic states of the K valley. Thus, we can start our discussions from the following continuum model.}
\begin{equation}
H_{\xi}(\boldsymbol{r}) = \left( \begin{array}{cc}
h_{b}(\boldsymbol{r}) & \hat{T}(\boldsymbol{r})(1+e^{ik_z}) \\
\hat{T}^{\dagger}(\boldsymbol{r})(1+e^{-ik_z}) & h_{t}(\boldsymbol{r})
\end{array} \right)
\end{equation}
\textcolor{black}{Here $h_{b}$ and $h_{t}$ represent the effective Hamiltonian of the monolayer near the $\xi$K valley. $\hat{T}$ is the interlayer tunneling term, and $e^{ik_z}$ originates from the interlayer coupling between neighboring unitcells in the twisted bulk. The detailed parameters in this effective model follow the previous work\cite{PhysRevLett.122.086402,PhysRevLett.131.016001}. Our calculated energy bands exhibit the $k_z$-dependent electronic structure as shown in the Fig.\,6(a-b).} 

\textcolor{black}{We consider the pairing matrices of interlayer triplet channel $\gamma _{y0}=\tau_y$ and $\gamma _{yz} = \tau_ys_z$. The pairing susceptibility with $\mathbf{q}$-finite-momentum pairing is described as}
\begin{equation}
    \chi_{ij,\mu\nu}(\mathbf{q}) = \int_{\mathbf{p}} \sum_{a,b} O_{a,b}^{j\nu}(\mathbf{p},\mathbf{q}) O_{a,b}^{i\mu*}(\mathbf{p},\mathbf{q}) \mathcal{K}_{ab}(\mathbf{p},\mathbf{q})
\end{equation}
\textcolor{black}{The momenta of electron and hole in the Cooper pair are $\mathbf{p}$ and $-\mathbf{p}$, respectively. The overlap function is defined as $O_{a,b}^{j\nu}(\mathbf{p},\mathbf{q}) = \langle u_{a\mathbf{p}+\mathbf{q}/2} | \gamma_{j\nu} | v_{b-\mathbf{p}+\mathbf{q}/2} \rangle$ where $a$ and $b$ are band indices. $|u\rangle$ and $|v\rangle$ are the eigenstates of electron and hole respectively, and the kernel function is }
\begin{equation}
    \mathcal{K}_{ab}(\pmb{p},\pmb{q}) = \frac{1 - f(E_{a}(\pmb{p} + \pmb{q}/2)) - f(E_{b}^{\prime}(-\pmb{p} + \pmb{q}/2))}{E_{a}(\pmb{p} + \pmb{q}/2) + E_{b}^{\prime}(-\pmb{p} + \pmb{q}/2)}
\end{equation}
\textcolor{black}{Here $f$ is the Fermi distribution function. $E_a$ and $E_b'$ are the eigenvalues of electron and hole, respectively. In this work, we focus on the superconductivity pairing with zero-momentum pairing ($\mathbf{q}=0$). We calculated the $k_z$-resolved pairing susceptibility defined as}
\begin{widetext}
\begin{equation}
    \chi_{ij,\mu\nu}^{k_z}(\mathbf{q}=0) = \int_{k_xk_y} \sum_{a,b} O_{a,b}^{j\nu}(k_x,k_y,k_z,\mathbf{q}=0) O_{a,b}^{i\mu*}(k_x,k_y,k_z,\mathbf{q}=0) \mathcal{K}_{ab}(k_x,k_y,k_z,\mathbf{q}=0).
\end{equation}
\end{widetext}
\textcolor{black}{Our calculated $\chi_{ij,\mu\nu}^{k_z}$ at different chemical potentials are shown in the Fig.\,6(c-d). The calculated results show that the non-zero $k_z$ term provides much larger pairing susceptibility compared to $k_z=0$-resolved pairing susceptibility, which indicates the important role of non-zero $k_z$ pairing susceptibility. Our results about effective Hamiltonian and pairing susceptibility show that electronic structures at non-zero $k_z$ planes remarkably amplify the interlayer triplet pairing in the twisted moiré bulk.}

\subsection{More discussions about triplet pairing superconductivity}
\textcolor{black}{A significant volume of research has been devoted to the three-dimensional spiral structure of transition metal dichalcogenides (TMD) associated with screw dislocations. This structure has been successfully grown through chemical vapor deposition (CVD) experiments. Consequently, the alternating twisted bulk TMD can be synthesized by means of similar experimental techniques.}

The minimal structural platform exhibiting the crucial mirror symmetry discussed here would be the alternating twisted trilayer system. For instance, alternating twisted trilayer graphene has been proposed as a promising platform for realizing spin-triplet pairing superconductivity, largely attributed to its mirror symmetry\cite{PhysRevLett.127.097001}. Similarly, alternating twisted trilayer NbSe$_2$ would also possess mirror symmetry and could potentially support triplet pairing. However, a significant experimental consideration is that such trilayer systems are inherently sensitive to substrate effects, which typically break the mirror symmetry. In contrast, the bulk nature of the alternating twisted moir\'e system investigated in this work offers inherent protection for its bulk states against such symmetry-breaking substrate potentials, potentially providing a more robust environment for realizing symmetry-protected phenomena.

\textcolor{black}{We now shift our focus to discussions regarding crucial spectroscopic and transport measurements that can offer characteristic signatures of interlayer triplet pairing. (i) Scanning tunneling microscopy or spectroscopy measures the local density of states, furnishing direct information on the superconducting gap in real-space. (ii) Additionally, nuclear magnetic resonance technology can assist in distinguishing triplet from singlet pairing. In a conventional spin-singlet superconductor, the spin susceptibility declines below the critical temperature T$_c$, resulting in a decrease in the Knight shift and a characteristic peak (the Hebel-Slichter peak) in the spin-lattice relaxation rate just below T$_c$, followed by an exponential decay at lower temperatures. Conversely, for spin-triplet pairing, the Knight shift remains invariant below Tc when a magnetic field is applied, as spin polarization is maintained in the Cooper pairs. Simultaneously, the spin-lattice relaxation rate will not display a Hebel-Slichter peak but will instead follow a power-law behavior at low temperatures. (iii) The fabrication of Josephson junctions, either vertically with the moiré bulk between two conventional s-wave superconductors (e.g., Nb) or in a planar geometry, can yield critical insights. The critical current in such junctions is sensitive to the phase and symmetry of the order parameter. A characteristic feature of triplet pairing is the potential for a Josephson current that exhibits a distinct dependence on the orientation of an in-plane magnetic field. Moreover, phase-sensitive measurements, such as those employing SQUID interferometers, could, in principle, detect an anomalous phase shift if the order parameter has an odd-parity component.}

\textcolor{black}{To probe the \textit{bulk} superconducting gap, more bulk-sensitive techniques such as soft X-ray ARPES or hard X-ray photoemission spectroscopy (HAXPES) would be required. The triplet superconductivity can also be detected through transport measurements, because the bulk signal is much stronger than the surface signal. The surface parallel channel might cause a broadening of the resistive transition or a residual resistance, but the emergence of a zero-resistance state remains a definitive hallmark of bulk superconductivity. To disentangle the contributions, standard van der Pauw measurements on samples of varying thickness can be employed, where a bulk-dominated signal should become thickness-independent beyond a certain scale.}

\textcolor{black}{The parent compound NbSe$_2$ exhibits the charge density wave (CDW) order due to electron-phonon interaction. In the CDW state, time-reversal, $C_{3z}$, $C_{2y}$, and mirror symmetries are still preserved, which maintain the interlayer triplet superconductivity in the twisted bulk NbSe$_2$ with CDW charge order. However, the CDW order may affect the transition temperature of interlayer triplet superconductivity. The interaction between CDW and interlayer triplet superconductivity remains an area for future theoretical and experimental investigations.}

\textcolor{black}{In the above discussions, we only address the pairing mechanism in the non-interacting case. In the twisted bulk NbSe$_2$, electron correlations may drive the system into distinct correlated states. For instance, the valley-polarized state that breaks the time-reversal symmetry or the nematic state that breaks the $C_{3z}$ symmetry eliminates the constraints on pairings of one-dimensional irreducible representation. Consequently, electron interaction may lead to a mixture of interlayer triplet pairing and other possible pairings, and the dressed electron-phonon interaction further impacts the robustness of the superconducting phase against other charge orders.} 

\section{Conclusion}
In summary, we studied three-dimensional flat bands in the alternating twisted NbSe$_2$ moir\'e bulk by first-principles calculations. Unlike the twisted bilayer system, each layer in the moir\'e bulk experiences stronger interlayer interactions. Our calculations reveal that the moir'e bulk undergoes spontaneous large-scale structural relaxation and exhibits remarkably flat energy bands when twist angle is not larger than 7.31°. The flat bands in various moir\'e bulks behave $k_z$-dependent, indicate of three-dimensional flat bands in the alternating twisted NbSe$_2$ moir\'e bulk. The out-of-plane mirror symmetry in the moir\'e bulk make its unique interlayer triplet pairing of superconductivity different from twisted bilayer system.

\section{acknowledgments}
This work was supported by the National Key Research and Development Program of China (No. 2024YFA1410300), the National Natural Science Foundation of China (No. 12304217), the Natural Science Foundation of Hunan Province (No. 2025JJ60002) and the Fundamental Research Funds for the Central Universities from China (No. 531119200247). We gratefully acknowledge HZWTECH for providing computation facilities.


%


\begin{thebibliography}{86}%
\makeatletter
\providecommand \@ifxundefined [1]{%
 \@ifx{#1\undefined}
}%
\providecommand \@ifnum [1]{%
 \ifnum #1\expandafter \@firstoftwo
 \else \expandafter \@secondoftwo
 \fi
}%
\providecommand \@ifx [1]{%
 \ifx #1\expandafter \@firstoftwo
 \else \expandafter \@secondoftwo
 \fi
}%
\providecommand \natexlab [1]{#1}%
\providecommand \enquote  [1]{``#1''}%
\providecommand \bibnamefont  [1]{#1}%
\providecommand \bibfnamefont [1]{#1}%
\providecommand \citenamefont [1]{#1}%
\providecommand \href@noop [0]{\@secondoftwo}%
\providecommand \href [0]{\begingroup \@sanitize@url \@href}%
\providecommand \@href[1]{\@@startlink{#1}\@@href}%
\providecommand \@@href[1]{\endgroup#1\@@endlink}%
\providecommand \@sanitize@url [0]{\catcode `\\12\catcode `\$12\catcode `\&12\catcode `\#12\catcode `\^12\catcode `\_12\catcode `\%12\relax}%
\providecommand \@@startlink[1]{}%
\providecommand \@@endlink[0]{}%
\providecommand \url  [0]{\begingroup\@sanitize@url \@url }%
\providecommand \@url [1]{\endgroup\@href {#1}{\urlprefix }}%
\providecommand \urlprefix  [0]{URL }%
\providecommand \Eprint [0]{\href }%
\providecommand \doibase [0]{https://doi.org/}%
\providecommand \selectlanguage [0]{\@gobble}%
\providecommand \bibinfo  [0]{\@secondoftwo}%
\providecommand \bibfield  [0]{\@secondoftwo}%
\providecommand \translation [1]{[#1]}%
\providecommand \BibitemOpen [0]{}%
\providecommand \bibitemStop [0]{}%
\providecommand \bibitemNoStop [0]{.\EOS\space}%
\providecommand \EOS [0]{\spacefactor3000\relax}%
\providecommand \BibitemShut  [1]{\csname bibitem#1\endcsname}%
\let\auto@bib@innerbib\@empty
\bibitem [{\citenamefont {Aoki}(2020)}]{SC1}%
  \BibitemOpen
  \bibfield  {author} {\bibinfo {author} {\bibfnamefont {H.}~\bibnamefont {Aoki}},\ }\href {https://doi.org/10.1007/s10948-020-05474-6} {\bibfield  {journal} {\bibinfo  {journal} {Journal of Superconductivity and Novel Magnetism}\ }\textbf {\bibinfo {volume} {33}},\ \bibinfo {pages} {2341} (\bibinfo {year} {2020})}\BibitemShut {NoStop}%
\bibitem [{\citenamefont {Cao}\ \emph {et~al.}(2018{\natexlab{a}})\citenamefont {Cao}, \citenamefont {Fatemi}, \citenamefont {Fang}, \citenamefont {Watanabe}, \citenamefont {Taniguchi}, \citenamefont {Kaxiras},\ and\ \citenamefont {Jarillo-Herrero}}]{TBG1}%
  \BibitemOpen
  \bibfield  {author} {\bibinfo {author} {\bibfnamefont {Y.}~\bibnamefont {Cao}}, \bibinfo {author} {\bibfnamefont {V.}~\bibnamefont {Fatemi}}, \bibinfo {author} {\bibfnamefont {S.}~\bibnamefont {Fang}}, \bibinfo {author} {\bibfnamefont {K.}~\bibnamefont {Watanabe}}, \bibinfo {author} {\bibfnamefont {T.}~\bibnamefont {Taniguchi}}, \bibinfo {author} {\bibfnamefont {E.}~\bibnamefont {Kaxiras}},\ and\ \bibinfo {author} {\bibfnamefont {P.}~\bibnamefont {Jarillo-Herrero}},\ }\href {https://api.semanticscholar.org/CorpusID:4655887} {\bibfield  {journal} {\bibinfo  {journal} {Nature}\ }\textbf {\bibinfo {volume} {556}},\ \bibinfo {pages} {43} (\bibinfo {year} {2018}{\natexlab{a}})}\BibitemShut {NoStop}%
\bibitem [{\citenamefont {Cao}\ \emph {et~al.}(2018{\natexlab{b}})\citenamefont {Cao}, \citenamefont {Fatemi}, \citenamefont {Demir}, \citenamefont {Fang}, \citenamefont {Tomarken}, \citenamefont {Luo}, \citenamefont {Sanchez-Yamagishi}, \citenamefont {Watanabe}, \citenamefont {Taniguchi}, \citenamefont {Kaxiras}, \citenamefont {Ashoori},\ and\ \citenamefont {Jarillo-Herrero}}]{TBG2}%
  \BibitemOpen
  \bibfield  {author} {\bibinfo {author} {\bibfnamefont {Y.}~\bibnamefont {Cao}}, \bibinfo {author} {\bibfnamefont {V.}~\bibnamefont {Fatemi}}, \bibinfo {author} {\bibfnamefont {A.}~\bibnamefont {Demir}}, \bibinfo {author} {\bibfnamefont {S.}~\bibnamefont {Fang}}, \bibinfo {author} {\bibfnamefont {S.~L.}\ \bibnamefont {Tomarken}}, \bibinfo {author} {\bibfnamefont {J.~Y.}\ \bibnamefont {Luo}}, \bibinfo {author} {\bibfnamefont {J.~D.}\ \bibnamefont {Sanchez-Yamagishi}}, \bibinfo {author} {\bibfnamefont {K.}~\bibnamefont {Watanabe}}, \bibinfo {author} {\bibfnamefont {T.}~\bibnamefont {Taniguchi}}, \bibinfo {author} {\bibfnamefont {E.}~\bibnamefont {Kaxiras}}, \bibinfo {author} {\bibfnamefont {R.~C.}\ \bibnamefont {Ashoori}},\ and\ \bibinfo {author} {\bibfnamefont {P.}~\bibnamefont {Jarillo-Herrero}},\ }\href {https://doi.org/10.1038/nature26154} {\bibfield  {journal} {\bibinfo  {journal} {Nature}\ }\textbf {\bibinfo {volume} {556}},\ \bibinfo {pages} {80} (\bibinfo {year} {2018}{\natexlab{b}})}\BibitemShut
  {NoStop}%
\bibitem [{\citenamefont {Tang}\ \emph {et~al.}(2020)\citenamefont {Tang}, \citenamefont {Li}, \citenamefont {Li}, \citenamefont {Xu}, \citenamefont {Liu}, \citenamefont {Barmak}, \citenamefont {Watanabe}, \citenamefont {Taniguchi}, \citenamefont {MacDonald}, \citenamefont {Shan},\ and\ \citenamefont {Mak}}]{flat1}%
  \BibitemOpen
  \bibfield  {author} {\bibinfo {author} {\bibfnamefont {Y.}~\bibnamefont {Tang}}, \bibinfo {author} {\bibfnamefont {L.}~\bibnamefont {Li}}, \bibinfo {author} {\bibfnamefont {T.}~\bibnamefont {Li}}, \bibinfo {author} {\bibfnamefont {Y.}~\bibnamefont {Xu}}, \bibinfo {author} {\bibfnamefont {S.}~\bibnamefont {Liu}}, \bibinfo {author} {\bibfnamefont {K.}~\bibnamefont {Barmak}}, \bibinfo {author} {\bibfnamefont {K.}~\bibnamefont {Watanabe}}, \bibinfo {author} {\bibfnamefont {T.}~\bibnamefont {Taniguchi}}, \bibinfo {author} {\bibfnamefont {A.~H.}\ \bibnamefont {MacDonald}}, \bibinfo {author} {\bibfnamefont {J.}~\bibnamefont {Shan}},\ and\ \bibinfo {author} {\bibfnamefont {K.~F.}\ \bibnamefont {Mak}},\ }\href {https://doi.org/10.1038/s41586-020-2085-3} {\bibfield  {journal} {\bibinfo  {journal} {Nature}\ }\textbf {\bibinfo {volume} {579}},\ \bibinfo {pages} {353} (\bibinfo {year} {2020})}\BibitemShut {NoStop}%
\bibitem [{\citenamefont {Regan}\ \emph {et~al.}(2020)\citenamefont {Regan}, \citenamefont {Wang}, \citenamefont {Jin}, \citenamefont {Bakti~Utama}, \citenamefont {Gao}, \citenamefont {Wei}, \citenamefont {Zhao}, \citenamefont {Zhao}, \citenamefont {Zhang}, \citenamefont {Yumigeta}, \citenamefont {Blei}, \citenamefont {Carlström}, \citenamefont {Watanabe}, \citenamefont {Taniguchi}, \citenamefont {Tongay}, \citenamefont {Crommie}, \citenamefont {Zettl},\ and\ \citenamefont {Wang}}]{flat2}%
  \BibitemOpen
  \bibfield  {author} {\bibinfo {author} {\bibfnamefont {E.~C.}\ \bibnamefont {Regan}}, \bibinfo {author} {\bibfnamefont {D.}~\bibnamefont {Wang}}, \bibinfo {author} {\bibfnamefont {C.}~\bibnamefont {Jin}}, \bibinfo {author} {\bibfnamefont {M.~I.}\ \bibnamefont {Bakti~Utama}}, \bibinfo {author} {\bibfnamefont {B.}~\bibnamefont {Gao}}, \bibinfo {author} {\bibfnamefont {X.}~\bibnamefont {Wei}}, \bibinfo {author} {\bibfnamefont {S.}~\bibnamefont {Zhao}}, \bibinfo {author} {\bibfnamefont {W.}~\bibnamefont {Zhao}}, \bibinfo {author} {\bibfnamefont {Z.}~\bibnamefont {Zhang}}, \bibinfo {author} {\bibfnamefont {K.}~\bibnamefont {Yumigeta}}, \bibinfo {author} {\bibfnamefont {M.}~\bibnamefont {Blei}}, \bibinfo {author} {\bibfnamefont {J.~D.}\ \bibnamefont {Carlström}}, \bibinfo {author} {\bibfnamefont {K.}~\bibnamefont {Watanabe}}, \bibinfo {author} {\bibfnamefont {T.}~\bibnamefont {Taniguchi}}, \bibinfo {author} {\bibfnamefont {S.}~\bibnamefont {Tongay}}, \bibinfo {author} {\bibfnamefont {M.}~\bibnamefont {Crommie}},
  \bibinfo {author} {\bibfnamefont {A.}~\bibnamefont {Zettl}},\ and\ \bibinfo {author} {\bibfnamefont {F.}~\bibnamefont {Wang}},\ }\href {https://doi.org/10.1038/s41586-020-2092-4} {\bibfield  {journal} {\bibinfo  {journal} {Nature}\ }\textbf {\bibinfo {volume} {579}},\ \bibinfo {pages} {359} (\bibinfo {year} {2020})}\BibitemShut {NoStop}%
\bibitem [{\citenamefont {Wang}\ \emph {et~al.}(2020)\citenamefont {Wang}, \citenamefont {Shih}, \citenamefont {Ghiotto}, \citenamefont {Xian}, \citenamefont {Rhodes}, \citenamefont {Tan}, \citenamefont {Claassen}, \citenamefont {Kennes}, \citenamefont {Bai}, \citenamefont {Kim}, \citenamefont {Watanabe}, \citenamefont {Taniguchi}, \citenamefont {Zhu}, \citenamefont {Hone}, \citenamefont {Rubio}, \citenamefont {Pasupathy},\ and\ \citenamefont {Dean}}]{flat3}%
  \BibitemOpen
  \bibfield  {author} {\bibinfo {author} {\bibfnamefont {L.}~\bibnamefont {Wang}}, \bibinfo {author} {\bibfnamefont {E.-M.}\ \bibnamefont {Shih}}, \bibinfo {author} {\bibfnamefont {A.}~\bibnamefont {Ghiotto}}, \bibinfo {author} {\bibfnamefont {L.}~\bibnamefont {Xian}}, \bibinfo {author} {\bibfnamefont {D.~A.}\ \bibnamefont {Rhodes}}, \bibinfo {author} {\bibfnamefont {C.}~\bibnamefont {Tan}}, \bibinfo {author} {\bibfnamefont {M.}~\bibnamefont {Claassen}}, \bibinfo {author} {\bibfnamefont {D.~M.}\ \bibnamefont {Kennes}}, \bibinfo {author} {\bibfnamefont {Y.}~\bibnamefont {Bai}}, \bibinfo {author} {\bibfnamefont {B.}~\bibnamefont {Kim}}, \bibinfo {author} {\bibfnamefont {K.}~\bibnamefont {Watanabe}}, \bibinfo {author} {\bibfnamefont {T.}~\bibnamefont {Taniguchi}}, \bibinfo {author} {\bibfnamefont {X.}~\bibnamefont {Zhu}}, \bibinfo {author} {\bibfnamefont {J.}~\bibnamefont {Hone}}, \bibinfo {author} {\bibfnamefont {A.}~\bibnamefont {Rubio}}, \bibinfo {author} {\bibfnamefont {A.~N.}\ \bibnamefont {Pasupathy}},\
  and\ \bibinfo {author} {\bibfnamefont {C.~R.}\ \bibnamefont {Dean}},\ }\href {https://doi.org/10.1038/s41563-020-0708-6} {\bibfield  {journal} {\bibinfo  {journal} {Nature Materials}\ }\textbf {\bibinfo {volume} {19}},\ \bibinfo {pages} {861} (\bibinfo {year} {2020})}\BibitemShut {NoStop}%
\bibitem [{\citenamefont {Pasquale}\ \emph {et~al.}(2022)\citenamefont {Pasquale}, \citenamefont {Sun}, \citenamefont {Čerņevičs}, \citenamefont {Perea-Causin}, \citenamefont {Tagarelli}, \citenamefont {Watanabe}, \citenamefont {Taniguchi}, \citenamefont {Malic}, \citenamefont {Yazyev},\ and\ \citenamefont {Kis}}]{flat4}%
  \BibitemOpen
  \bibfield  {author} {\bibinfo {author} {\bibfnamefont {G.}~\bibnamefont {Pasquale}}, \bibinfo {author} {\bibfnamefont {Z.}~\bibnamefont {Sun}}, \bibinfo {author} {\bibfnamefont {K.}~\bibnamefont {Čerņevičs}}, \bibinfo {author} {\bibfnamefont {R.}~\bibnamefont {Perea-Causin}}, \bibinfo {author} {\bibfnamefont {F.}~\bibnamefont {Tagarelli}}, \bibinfo {author} {\bibfnamefont {K.}~\bibnamefont {Watanabe}}, \bibinfo {author} {\bibfnamefont {T.}~\bibnamefont {Taniguchi}}, \bibinfo {author} {\bibfnamefont {E.}~\bibnamefont {Malic}}, \bibinfo {author} {\bibfnamefont {O.~V.}\ \bibnamefont {Yazyev}},\ and\ \bibinfo {author} {\bibfnamefont {A.}~\bibnamefont {Kis}},\ }\href {https://doi.org/10.1021/acs.nanolett.2c02965} {\bibfield  {journal} {\bibinfo  {journal} {Nano Letters}\ }\textbf {\bibinfo {volume} {22}},\ \bibinfo {pages} {8883} (\bibinfo {year} {2022})}\BibitemShut {NoStop}%
\bibitem [{\citenamefont {Wilhelm}\ \emph {et~al.}(2021)\citenamefont {Wilhelm}, \citenamefont {Lang},\ and\ \citenamefont {Läuchli}}]{FCI}%
  \BibitemOpen
  \bibfield  {author} {\bibinfo {author} {\bibfnamefont {P.}~\bibnamefont {Wilhelm}}, \bibinfo {author} {\bibfnamefont {T.~C.}\ \bibnamefont {Lang}},\ and\ \bibinfo {author} {\bibfnamefont {A.~M.}\ \bibnamefont {Läuchli}},\ }\bibfield  {journal} {\bibinfo  {journal} {Physical review. B./Physical review. B}\ }\textbf {\bibinfo {volume} {103}},\ \href {https://doi.org/10.1103/physrevb.103.125406} {10.1103/physrevb.103.125406} (\bibinfo {year} {2021})\BibitemShut {NoStop}%
\bibitem [{\citenamefont {Zhang}\ \emph {et~al.}(2005)\citenamefont {Zhang}, \citenamefont {Tan}, \citenamefont {Stormer},\ and\ \citenamefont {Kim}}]{QHE}%
  \BibitemOpen
  \bibfield  {author} {\bibinfo {author} {\bibfnamefont {Y.}~\bibnamefont {Zhang}}, \bibinfo {author} {\bibfnamefont {Y.-W.}\ \bibnamefont {Tan}}, \bibinfo {author} {\bibfnamefont {H.~L.}\ \bibnamefont {Stormer}},\ and\ \bibinfo {author} {\bibfnamefont {P.}~\bibnamefont {Kim}},\ }\href {https://doi.org/10.1038/nature04235} {\bibfield  {journal} {\bibinfo  {journal} {Nature}\ }\textbf {\bibinfo {volume} {438}},\ \bibinfo {pages} {201–204} (\bibinfo {year} {2005})}\BibitemShut {NoStop}%
\bibitem [{\citenamefont {T{\"o}rm{\"a}}\ \emph {et~al.}(2022)\citenamefont {T{\"o}rm{\"a}}, \citenamefont {Peotta},\ and\ \citenamefont {Bernevig}}]{torma2022superconductivity}%
  \BibitemOpen
  \bibfield  {author} {\bibinfo {author} {\bibfnamefont {P.}~\bibnamefont {T{\"o}rm{\"a}}}, \bibinfo {author} {\bibfnamefont {S.}~\bibnamefont {Peotta}},\ and\ \bibinfo {author} {\bibfnamefont {B.~A.}\ \bibnamefont {Bernevig}},\ }\href@noop {} {\bibfield  {journal} {\bibinfo  {journal} {Nature Reviews Physics}\ }\textbf {\bibinfo {volume} {4}},\ \bibinfo {pages} {528} (\bibinfo {year} {2022})}\BibitemShut {NoStop}%
\bibitem [{\citenamefont {Lee}\ \emph {et~al.}(2019)\citenamefont {Lee}, \citenamefont {Khalaf}, \citenamefont {Liu}, \citenamefont {Liu}, \citenamefont {Hao}, \citenamefont {Kim},\ and\ \citenamefont {Vishwanath}}]{lee2019theory}%
  \BibitemOpen
  \bibfield  {author} {\bibinfo {author} {\bibfnamefont {J.~Y.}\ \bibnamefont {Lee}}, \bibinfo {author} {\bibfnamefont {E.}~\bibnamefont {Khalaf}}, \bibinfo {author} {\bibfnamefont {S.}~\bibnamefont {Liu}}, \bibinfo {author} {\bibfnamefont {X.}~\bibnamefont {Liu}}, \bibinfo {author} {\bibfnamefont {Z.}~\bibnamefont {Hao}}, \bibinfo {author} {\bibfnamefont {P.}~\bibnamefont {Kim}},\ and\ \bibinfo {author} {\bibfnamefont {A.}~\bibnamefont {Vishwanath}},\ }\href@noop {} {\bibfield  {journal} {\bibinfo  {journal} {Nature communications}\ }\textbf {\bibinfo {volume} {10}},\ \bibinfo {pages} {5333} (\bibinfo {year} {2019})}\BibitemShut {NoStop}%
\bibitem [{\citenamefont {Wu}\ \emph {et~al.}(2018)\citenamefont {Wu}, \citenamefont {MacDonald},\ and\ \citenamefont {Martin}}]{wu2018theory}%
  \BibitemOpen
  \bibfield  {author} {\bibinfo {author} {\bibfnamefont {F.}~\bibnamefont {Wu}}, \bibinfo {author} {\bibfnamefont {A.~H.}\ \bibnamefont {MacDonald}},\ and\ \bibinfo {author} {\bibfnamefont {I.}~\bibnamefont {Martin}},\ }\href@noop {} {\bibfield  {journal} {\bibinfo  {journal} {Physical review letters}\ }\textbf {\bibinfo {volume} {121}},\ \bibinfo {pages} {257001} (\bibinfo {year} {2018})}\BibitemShut {NoStop}%
\bibitem [{\citenamefont {Codecido}\ \emph {et~al.}(2019)\citenamefont {Codecido}, \citenamefont {Wang}, \citenamefont {Koester}, \citenamefont {Che}, \citenamefont {Tian}, \citenamefont {Lv}, \citenamefont {Tran}, \citenamefont {Watanabe}, \citenamefont {Taniguchi}, \citenamefont {Zhang}, \citenamefont {Bockrath},\ and\ \citenamefont {Lau}}]{marc-tbg-19}%
  \BibitemOpen
  \bibfield  {author} {\bibinfo {author} {\bibfnamefont {E.}~\bibnamefont {Codecido}}, \bibinfo {author} {\bibfnamefont {Q.}~\bibnamefont {Wang}}, \bibinfo {author} {\bibfnamefont {R.}~\bibnamefont {Koester}}, \bibinfo {author} {\bibfnamefont {S.}~\bibnamefont {Che}}, \bibinfo {author} {\bibfnamefont {H.}~\bibnamefont {Tian}}, \bibinfo {author} {\bibfnamefont {R.}~\bibnamefont {Lv}}, \bibinfo {author} {\bibfnamefont {S.}~\bibnamefont {Tran}}, \bibinfo {author} {\bibfnamefont {K.}~\bibnamefont {Watanabe}}, \bibinfo {author} {\bibfnamefont {T.}~\bibnamefont {Taniguchi}}, \bibinfo {author} {\bibfnamefont {F.}~\bibnamefont {Zhang}}, \bibinfo {author} {\bibfnamefont {M.}~\bibnamefont {Bockrath}},\ and\ \bibinfo {author} {\bibfnamefont {C.~N.}\ \bibnamefont {Lau}},\ }\bibfield  {journal} {\bibinfo  {journal} {Science Advances}\ }\textbf {\bibinfo {volume} {5}},\ \href {https://doi.org/10.1126/sciadv.aaw9770} {10.1126/sciadv.aaw9770} (\bibinfo {year} {2019})\BibitemShut {NoStop}%
\bibitem [{\citenamefont {Lu}\ \emph {et~al.}(2019)\citenamefont {Lu}, \citenamefont {Stepanov}, \citenamefont {Yang}, \citenamefont {Xie}, \citenamefont {Aamir}, \citenamefont {Das}, \citenamefont {Urgell}, \citenamefont {Watanabe}, \citenamefont {Taniguchi}, \citenamefont {Zhang}, \citenamefont {Bachtold}, \citenamefont {MacDonald},\ and\ \citenamefont {Efetov}}]{efetov-nature19}%
  \BibitemOpen
  \bibfield  {author} {\bibinfo {author} {\bibfnamefont {X.}~\bibnamefont {Lu}}, \bibinfo {author} {\bibfnamefont {P.}~\bibnamefont {Stepanov}}, \bibinfo {author} {\bibfnamefont {W.}~\bibnamefont {Yang}}, \bibinfo {author} {\bibfnamefont {M.}~\bibnamefont {Xie}}, \bibinfo {author} {\bibfnamefont {M.~A.}\ \bibnamefont {Aamir}}, \bibinfo {author} {\bibfnamefont {I.}~\bibnamefont {Das}}, \bibinfo {author} {\bibfnamefont {C.}~\bibnamefont {Urgell}}, \bibinfo {author} {\bibfnamefont {K.}~\bibnamefont {Watanabe}}, \bibinfo {author} {\bibfnamefont {T.}~\bibnamefont {Taniguchi}}, \bibinfo {author} {\bibfnamefont {G.}~\bibnamefont {Zhang}}, \bibinfo {author} {\bibfnamefont {A.}~\bibnamefont {Bachtold}}, \bibinfo {author} {\bibfnamefont {A.~H.}\ \bibnamefont {MacDonald}},\ and\ \bibinfo {author} {\bibfnamefont {D.~K.}\ \bibnamefont {Efetov}},\ }\href {https://doi.org/10.1038/s41586-019-1695-0} {\bibfield  {journal} {\bibinfo  {journal} {Nature}\ }\textbf {\bibinfo {volume} {574}},\ \bibinfo {pages} {653} (\bibinfo
  {year} {2019})}\BibitemShut {NoStop}%
\bibitem [{\citenamefont {Stepanov}\ \emph {et~al.}(2020)\citenamefont {Stepanov}, \citenamefont {Das}, \citenamefont {Lu}, \citenamefont {Fahimniya}, \citenamefont {Watanabe}, \citenamefont {Taniguchi}, \citenamefont {Koppens}, \citenamefont {Lischner}, \citenamefont {Levitov},\ and\ \citenamefont {Efetov}}]{efetov-nature20}%
  \BibitemOpen
  \bibfield  {author} {\bibinfo {author} {\bibfnamefont {P.}~\bibnamefont {Stepanov}}, \bibinfo {author} {\bibfnamefont {I.}~\bibnamefont {Das}}, \bibinfo {author} {\bibfnamefont {X.}~\bibnamefont {Lu}}, \bibinfo {author} {\bibfnamefont {A.}~\bibnamefont {Fahimniya}}, \bibinfo {author} {\bibfnamefont {K.}~\bibnamefont {Watanabe}}, \bibinfo {author} {\bibfnamefont {T.}~\bibnamefont {Taniguchi}}, \bibinfo {author} {\bibfnamefont {F.~H.~L.}\ \bibnamefont {Koppens}}, \bibinfo {author} {\bibfnamefont {J.}~\bibnamefont {Lischner}}, \bibinfo {author} {\bibfnamefont {L.}~\bibnamefont {Levitov}},\ and\ \bibinfo {author} {\bibfnamefont {D.~K.}\ \bibnamefont {Efetov}},\ }\href {https://doi.org/10.1038/s41586-020-2459-6} {\bibfield  {journal} {\bibinfo  {journal} {Nature}\ }\textbf {\bibinfo {volume} {583}},\ \bibinfo {pages} {375} (\bibinfo {year} {2020})}\BibitemShut {NoStop}%
\bibitem [{\citenamefont {Saito}\ \emph {et~al.}(2020)\citenamefont {Saito}, \citenamefont {Ge}, \citenamefont {Watanabe}, \citenamefont {Taniguchi},\ and\ \citenamefont {Young}}]{young-tbg-np20}%
  \BibitemOpen
  \bibfield  {author} {\bibinfo {author} {\bibfnamefont {Y.}~\bibnamefont {Saito}}, \bibinfo {author} {\bibfnamefont {J.}~\bibnamefont {Ge}}, \bibinfo {author} {\bibfnamefont {K.}~\bibnamefont {Watanabe}}, \bibinfo {author} {\bibfnamefont {T.}~\bibnamefont {Taniguchi}},\ and\ \bibinfo {author} {\bibfnamefont {A.~F.}\ \bibnamefont {Young}},\ }\href {https://doi.org/10.1038/s41567-020-0928-3} {\bibfield  {journal} {\bibinfo  {journal} {Nature Physics}\ }\textbf {\bibinfo {volume} {16}},\ \bibinfo {pages} {926} (\bibinfo {year} {2020})}\BibitemShut {NoStop}%
\bibitem [{\citenamefont {Liu}\ \emph {et~al.}(2021)\citenamefont {Liu}, \citenamefont {Wang}, \citenamefont {Watanabe}, \citenamefont {Taniguchi}, \citenamefont {Vafek},\ and\ \citenamefont {Li}}]{li-tbg-science21}%
  \BibitemOpen
  \bibfield  {author} {\bibinfo {author} {\bibfnamefont {X.}~\bibnamefont {Liu}}, \bibinfo {author} {\bibfnamefont {Z.}~\bibnamefont {Wang}}, \bibinfo {author} {\bibfnamefont {K.}~\bibnamefont {Watanabe}}, \bibinfo {author} {\bibfnamefont {T.}~\bibnamefont {Taniguchi}}, \bibinfo {author} {\bibfnamefont {O.}~\bibnamefont {Vafek}},\ and\ \bibinfo {author} {\bibfnamefont {J.}~\bibnamefont {Li}},\ }\href@noop {} {\bibfield  {journal} {\bibinfo  {journal} {Science}\ }\textbf {\bibinfo {volume} {371}},\ \bibinfo {pages} {1261} (\bibinfo {year} {2021})}\BibitemShut {NoStop}%
\bibitem [{\citenamefont {Cao}\ \emph {et~al.}(2021)\citenamefont {Cao}, \citenamefont {Rodan-Legrain}, \citenamefont {Park}, \citenamefont {Yuan}, \citenamefont {Watanabe}, \citenamefont {Taniguchi}, \citenamefont {Fernandes}, \citenamefont {Fu},\ and\ \citenamefont {Jarillo-Herrero}}]{cao-tbg-nematic-science21}%
  \BibitemOpen
  \bibfield  {author} {\bibinfo {author} {\bibfnamefont {Y.}~\bibnamefont {Cao}}, \bibinfo {author} {\bibfnamefont {D.}~\bibnamefont {Rodan-Legrain}}, \bibinfo {author} {\bibfnamefont {J.~M.}\ \bibnamefont {Park}}, \bibinfo {author} {\bibfnamefont {N.~F.}\ \bibnamefont {Yuan}}, \bibinfo {author} {\bibfnamefont {K.}~\bibnamefont {Watanabe}}, \bibinfo {author} {\bibfnamefont {T.}~\bibnamefont {Taniguchi}}, \bibinfo {author} {\bibfnamefont {R.~M.}\ \bibnamefont {Fernandes}}, \bibinfo {author} {\bibfnamefont {L.}~\bibnamefont {Fu}},\ and\ \bibinfo {author} {\bibfnamefont {P.}~\bibnamefont {Jarillo-Herrero}},\ }\href@noop {} {\bibfield  {journal} {\bibinfo  {journal} {science}\ }\textbf {\bibinfo {volume} {372}},\ \bibinfo {pages} {264} (\bibinfo {year} {2021})}\BibitemShut {NoStop}%
\bibitem [{\citenamefont {Kang}\ and\ \citenamefont {Vafek}(2019)}]{kang-tbg-prl19}%
  \BibitemOpen
  \bibfield  {author} {\bibinfo {author} {\bibfnamefont {J.}~\bibnamefont {Kang}}\ and\ \bibinfo {author} {\bibfnamefont {O.}~\bibnamefont {Vafek}},\ }\href@noop {} {\bibfield  {journal} {\bibinfo  {journal} {Phys. Rev. Lett.}\ }\textbf {\bibinfo {volume} {122}},\ \bibinfo {pages} {246401} (\bibinfo {year} {2019})}\BibitemShut {NoStop}%
\bibitem [{\citenamefont {Seo}\ \emph {et~al.}(2019)\citenamefont {Seo}, \citenamefont {Kotov},\ and\ \citenamefont {Uchoa}}]{Uchoa-ferroMott-prl}%
  \BibitemOpen
  \bibfield  {author} {\bibinfo {author} {\bibfnamefont {K.}~\bibnamefont {Seo}}, \bibinfo {author} {\bibfnamefont {V.~N.}\ \bibnamefont {Kotov}},\ and\ \bibinfo {author} {\bibfnamefont {B.}~\bibnamefont {Uchoa}},\ }\href {https://doi.org/10.1103/PhysRevLett.122.246402} {\bibfield  {journal} {\bibinfo  {journal} {Phys. Rev. Lett.}\ }\textbf {\bibinfo {volume} {122}},\ \bibinfo {pages} {246402} (\bibinfo {year} {2019})}\BibitemShut {NoStop}%
\bibitem [{\citenamefont {Xie}\ and\ \citenamefont {MacDonald}(2020)}]{xie-tbg-2018}%
  \BibitemOpen
  \bibfield  {author} {\bibinfo {author} {\bibfnamefont {M.}~\bibnamefont {Xie}}\ and\ \bibinfo {author} {\bibfnamefont {A.~H.}\ \bibnamefont {MacDonald}},\ }\href {https://doi.org/10.1103/PhysRevLett.124.097601} {\bibfield  {journal} {\bibinfo  {journal} {Phys. Rev. Lett.}\ }\textbf {\bibinfo {volume} {124}},\ \bibinfo {pages} {097601} (\bibinfo {year} {2020})}\BibitemShut {NoStop}%
\bibitem [{\citenamefont {Bultinck}\ \emph {et~al.}(2020{\natexlab{a}})\citenamefont {Bultinck}, \citenamefont {Chatterjee},\ and\ \citenamefont {Zaletel}}]{zaletel-tbg-2019}%
  \BibitemOpen
  \bibfield  {author} {\bibinfo {author} {\bibfnamefont {N.}~\bibnamefont {Bultinck}}, \bibinfo {author} {\bibfnamefont {S.}~\bibnamefont {Chatterjee}},\ and\ \bibinfo {author} {\bibfnamefont {M.~P.}\ \bibnamefont {Zaletel}},\ }\href {https://doi.org/10.1103/PhysRevLett.124.166601} {\bibfield  {journal} {\bibinfo  {journal} {Phys. Rev. Lett.}\ }\textbf {\bibinfo {volume} {124}},\ \bibinfo {pages} {166601} (\bibinfo {year} {2020}{\natexlab{a}})}\BibitemShut {NoStop}%
\bibitem [{\citenamefont {Zhang}\ \emph {et~al.}(2022{\natexlab{a}})\citenamefont {Zhang}, \citenamefont {Dai},\ and\ \citenamefont {Liu}}]{PhysRevLett.128.026403}%
  \BibitemOpen
  \bibfield  {author} {\bibinfo {author} {\bibfnamefont {S.}~\bibnamefont {Zhang}}, \bibinfo {author} {\bibfnamefont {X.}~\bibnamefont {Dai}},\ and\ \bibinfo {author} {\bibfnamefont {J.}~\bibnamefont {Liu}},\ }\href {https://doi.org/10.1103/PhysRevLett.128.026403} {\bibfield  {journal} {\bibinfo  {journal} {Phys. Rev. Lett.}\ }\textbf {\bibinfo {volume} {128}},\ \bibinfo {pages} {026403} (\bibinfo {year} {2022}{\natexlab{a}})}\BibitemShut {NoStop}%
\bibitem [{\citenamefont {Zhang}\ \emph {et~al.}(2022{\natexlab{b}})\citenamefont {Zhang}, \citenamefont {Lu},\ and\ \citenamefont {Liu}}]{PhysRevLett.128.247402}%
  \BibitemOpen
  \bibfield  {author} {\bibinfo {author} {\bibfnamefont {S.}~\bibnamefont {Zhang}}, \bibinfo {author} {\bibfnamefont {X.}~\bibnamefont {Lu}},\ and\ \bibinfo {author} {\bibfnamefont {J.}~\bibnamefont {Liu}},\ }\href {https://doi.org/10.1103/PhysRevLett.128.247402} {\bibfield  {journal} {\bibinfo  {journal} {Phys. Rev. Lett.}\ }\textbf {\bibinfo {volume} {128}},\ \bibinfo {pages} {247402} (\bibinfo {year} {2022}{\natexlab{b}})}\BibitemShut {NoStop}%
\bibitem [{\citenamefont {Bultinck}\ \emph {et~al.}(2020{\natexlab{b}})\citenamefont {Bultinck}, \citenamefont {Khalaf}, \citenamefont {Liu}, \citenamefont {Chatterjee}, \citenamefont {Vishwanath},\ and\ \citenamefont {Zaletel}}]{zaletel-tbg-hf-prx20}%
  \BibitemOpen
  \bibfield  {author} {\bibinfo {author} {\bibfnamefont {N.}~\bibnamefont {Bultinck}}, \bibinfo {author} {\bibfnamefont {E.}~\bibnamefont {Khalaf}}, \bibinfo {author} {\bibfnamefont {S.}~\bibnamefont {Liu}}, \bibinfo {author} {\bibfnamefont {S.}~\bibnamefont {Chatterjee}}, \bibinfo {author} {\bibfnamefont {A.}~\bibnamefont {Vishwanath}},\ and\ \bibinfo {author} {\bibfnamefont {M.~P.}\ \bibnamefont {Zaletel}},\ }\href {https://doi.org/10.1103/PhysRevX.10.031034} {\bibfield  {journal} {\bibinfo  {journal} {Phys. Rev. X}\ }\textbf {\bibinfo {volume} {10}},\ \bibinfo {pages} {031034} (\bibinfo {year} {2020}{\natexlab{b}})}\BibitemShut {NoStop}%
\bibitem [{\citenamefont {Liu}\ and\ \citenamefont {Dai}(2021)}]{jpliu-tbghf-prb21}%
  \BibitemOpen
  \bibfield  {author} {\bibinfo {author} {\bibfnamefont {J.}~\bibnamefont {Liu}}\ and\ \bibinfo {author} {\bibfnamefont {X.}~\bibnamefont {Dai}},\ }\href {https://doi.org/10.1103/PhysRevB.103.035427} {\bibfield  {journal} {\bibinfo  {journal} {Phys. Rev. B}\ }\textbf {\bibinfo {volume} {103}},\ \bibinfo {pages} {035427} (\bibinfo {year} {2021})}\BibitemShut {NoStop}%
\bibitem [{\citenamefont {Zhang}\ \emph {et~al.}(2020)\citenamefont {Zhang}, \citenamefont {Jiang}, \citenamefont {Wang},\ and\ \citenamefont {Zhang}}]{zhang-tbghf-arxiv20}%
  \BibitemOpen
  \bibfield  {author} {\bibinfo {author} {\bibfnamefont {Y.}~\bibnamefont {Zhang}}, \bibinfo {author} {\bibfnamefont {K.}~\bibnamefont {Jiang}}, \bibinfo {author} {\bibfnamefont {Z.}~\bibnamefont {Wang}},\ and\ \bibinfo {author} {\bibfnamefont {F.}~\bibnamefont {Zhang}},\ }\href {https://doi.org/10.1103/PhysRevB.102.035136} {\bibfield  {journal} {\bibinfo  {journal} {Phys. Rev. B}\ }\textbf {\bibinfo {volume} {102}},\ \bibinfo {pages} {035136} (\bibinfo {year} {2020})}\BibitemShut {NoStop}%
\bibitem [{\citenamefont {Hejazi}\ \emph {et~al.}(2021)\citenamefont {Hejazi}, \citenamefont {Chen},\ and\ \citenamefont {Balents}}]{hejazi-tbg-hf}%
  \BibitemOpen
  \bibfield  {author} {\bibinfo {author} {\bibfnamefont {K.}~\bibnamefont {Hejazi}}, \bibinfo {author} {\bibfnamefont {X.}~\bibnamefont {Chen}},\ and\ \bibinfo {author} {\bibfnamefont {L.}~\bibnamefont {Balents}},\ }\href {https://doi.org/10.1103/PhysRevResearch.3.013242} {\bibfield  {journal} {\bibinfo  {journal} {Phys. Rev. Research}\ }\textbf {\bibinfo {volume} {3}},\ \bibinfo {pages} {013242} (\bibinfo {year} {2021})}\BibitemShut {NoStop}%
\bibitem [{\citenamefont {Kang}\ and\ \citenamefont {Vafek}(2020)}]{kang-tbg-dmrg-prb20}%
  \BibitemOpen
  \bibfield  {author} {\bibinfo {author} {\bibfnamefont {J.}~\bibnamefont {Kang}}\ and\ \bibinfo {author} {\bibfnamefont {O.}~\bibnamefont {Vafek}},\ }\href {https://doi.org/10.1103/PhysRevB.102.035161} {\bibfield  {journal} {\bibinfo  {journal} {Phys. Rev. B}\ }\textbf {\bibinfo {volume} {102}},\ \bibinfo {pages} {035161} (\bibinfo {year} {2020})}\BibitemShut {NoStop}%
\bibitem [{\citenamefont {Chen}\ \emph {et~al.}(2021)\citenamefont {Chen}, \citenamefont {Liao}, \citenamefont {Chen}, \citenamefont {Vafek}, \citenamefont {Kang}, \citenamefont {Li},\ and\ \citenamefont {Meng}}]{kang-tbg-topomott}%
  \BibitemOpen
  \bibfield  {author} {\bibinfo {author} {\bibfnamefont {B.-B.}\ \bibnamefont {Chen}}, \bibinfo {author} {\bibfnamefont {Y.~D.}\ \bibnamefont {Liao}}, \bibinfo {author} {\bibfnamefont {Z.}~\bibnamefont {Chen}}, \bibinfo {author} {\bibfnamefont {O.}~\bibnamefont {Vafek}}, \bibinfo {author} {\bibfnamefont {J.}~\bibnamefont {Kang}}, \bibinfo {author} {\bibfnamefont {W.}~\bibnamefont {Li}},\ and\ \bibinfo {author} {\bibfnamefont {Z.~Y.}\ \bibnamefont {Meng}},\ }\href {https://doi.org/10.1038/s41467-021-25438-1} {\bibfield  {journal} {\bibinfo  {journal} {Nature Communications}\ }\textbf {\bibinfo {volume} {12}},\ \bibinfo {pages} {5480} (\bibinfo {year} {2021})}\BibitemShut {NoStop}%
\bibitem [{\citenamefont {Da~Liao}\ \emph {et~al.}(2021)\citenamefont {Da~Liao}, \citenamefont {Kang}, \citenamefont {Brei\o{}}, \citenamefont {Xu}, \citenamefont {Wu}, \citenamefont {Andersen}, \citenamefont {Fernandes},\ and\ \citenamefont {Meng}}]{meng-tbg-arxiv20}%
  \BibitemOpen
  \bibfield  {author} {\bibinfo {author} {\bibfnamefont {Y.}~\bibnamefont {Da~Liao}}, \bibinfo {author} {\bibfnamefont {J.}~\bibnamefont {Kang}}, \bibinfo {author} {\bibfnamefont {C.~N.}\ \bibnamefont {Brei\o{}}}, \bibinfo {author} {\bibfnamefont {X.~Y.}\ \bibnamefont {Xu}}, \bibinfo {author} {\bibfnamefont {H.-Q.}\ \bibnamefont {Wu}}, \bibinfo {author} {\bibfnamefont {B.~M.}\ \bibnamefont {Andersen}}, \bibinfo {author} {\bibfnamefont {R.~M.}\ \bibnamefont {Fernandes}},\ and\ \bibinfo {author} {\bibfnamefont {Z.~Y.}\ \bibnamefont {Meng}},\ }\href {https://doi.org/10.1103/PhysRevX.11.011014} {\bibfield  {journal} {\bibinfo  {journal} {Phys. Rev. X}\ }\textbf {\bibinfo {volume} {11}},\ \bibinfo {pages} {011014} (\bibinfo {year} {2021})}\BibitemShut {NoStop}%
\bibitem [{\citenamefont {Bernevig}\ \emph {et~al.}(2021)\citenamefont {Bernevig}, \citenamefont {Song}, \citenamefont {Regnault},\ and\ \citenamefont {Lian}}]{Bernevig-tbg3-arxiv20}%
  \BibitemOpen
  \bibfield  {author} {\bibinfo {author} {\bibfnamefont {B.~A.}\ \bibnamefont {Bernevig}}, \bibinfo {author} {\bibfnamefont {Z.-D.}\ \bibnamefont {Song}}, \bibinfo {author} {\bibfnamefont {N.}~\bibnamefont {Regnault}},\ and\ \bibinfo {author} {\bibfnamefont {B.}~\bibnamefont {Lian}},\ }\href {https://doi.org/10.1103/PhysRevB.103.205413} {\bibfield  {journal} {\bibinfo  {journal} {Phys. Rev. B}\ }\textbf {\bibinfo {volume} {103}},\ \bibinfo {pages} {205413} (\bibinfo {year} {2021})}\BibitemShut {NoStop}%
\bibitem [{\citenamefont {Lian}\ \emph {et~al.}(2021)\citenamefont {Lian}, \citenamefont {Song}, \citenamefont {Regnault}, \citenamefont {Efetov}, \citenamefont {Yazdani},\ and\ \citenamefont {Bernevig}}]{Lian-tbg4-arxiv20}%
  \BibitemOpen
  \bibfield  {author} {\bibinfo {author} {\bibfnamefont {B.}~\bibnamefont {Lian}}, \bibinfo {author} {\bibfnamefont {Z.-D.}\ \bibnamefont {Song}}, \bibinfo {author} {\bibfnamefont {N.}~\bibnamefont {Regnault}}, \bibinfo {author} {\bibfnamefont {D.~K.}\ \bibnamefont {Efetov}}, \bibinfo {author} {\bibfnamefont {A.}~\bibnamefont {Yazdani}},\ and\ \bibinfo {author} {\bibfnamefont {B.~A.}\ \bibnamefont {Bernevig}},\ }\href {https://doi.org/10.1103/PhysRevB.103.205414} {\bibfield  {journal} {\bibinfo  {journal} {Phys. Rev. B}\ }\textbf {\bibinfo {volume} {103}},\ \bibinfo {pages} {205414} (\bibinfo {year} {2021})}\BibitemShut {NoStop}%
\bibitem [{\citenamefont {Xie}\ \emph {et~al.}(2021)\citenamefont {Xie}, \citenamefont {Cowsik}, \citenamefont {Song}, \citenamefont {Lian}, \citenamefont {Bernevig},\ and\ \citenamefont {Regnault}}]{regnault-tbg-ed}%
  \BibitemOpen
  \bibfield  {author} {\bibinfo {author} {\bibfnamefont {F.}~\bibnamefont {Xie}}, \bibinfo {author} {\bibfnamefont {A.}~\bibnamefont {Cowsik}}, \bibinfo {author} {\bibfnamefont {Z.-D.}\ \bibnamefont {Song}}, \bibinfo {author} {\bibfnamefont {B.}~\bibnamefont {Lian}}, \bibinfo {author} {\bibfnamefont {B.~A.}\ \bibnamefont {Bernevig}},\ and\ \bibinfo {author} {\bibfnamefont {N.}~\bibnamefont {Regnault}},\ }\href {https://doi.org/10.1103/PhysRevB.103.205416} {\bibfield  {journal} {\bibinfo  {journal} {Phys. Rev. B}\ }\textbf {\bibinfo {volume} {103}},\ \bibinfo {pages} {205416} (\bibinfo {year} {2021})}\BibitemShut {NoStop}%
\bibitem [{\citenamefont {Soejima}\ \emph {et~al.}(2020)\citenamefont {Soejima}, \citenamefont {Parker}, \citenamefont {Bultinck}, \citenamefont {Hauschild},\ and\ \citenamefont {Zaletel}}]{zaletel-dmrg-prb20}%
  \BibitemOpen
  \bibfield  {author} {\bibinfo {author} {\bibfnamefont {T.}~\bibnamefont {Soejima}}, \bibinfo {author} {\bibfnamefont {D.~E.}\ \bibnamefont {Parker}}, \bibinfo {author} {\bibfnamefont {N.}~\bibnamefont {Bultinck}}, \bibinfo {author} {\bibfnamefont {J.}~\bibnamefont {Hauschild}},\ and\ \bibinfo {author} {\bibfnamefont {M.~P.}\ \bibnamefont {Zaletel}},\ }\href {https://doi.org/10.1103/PhysRevB.102.205111} {\bibfield  {journal} {\bibinfo  {journal} {Phys. Rev. B}\ }\textbf {\bibinfo {volume} {102}},\ \bibinfo {pages} {205111} (\bibinfo {year} {2020})}\BibitemShut {NoStop}%
\bibitem [{\citenamefont {Zhang}\ \emph {et~al.}(2021{\natexlab{a}})\citenamefont {Zhang}, \citenamefont {Pan}, \citenamefont {Zhang}, \citenamefont {Kang},\ and\ \citenamefont {Meng}}]{meng-tbg-qmc-cpl21}%
  \BibitemOpen
  \bibfield  {author} {\bibinfo {author} {\bibfnamefont {X.}~\bibnamefont {Zhang}}, \bibinfo {author} {\bibfnamefont {G.}~\bibnamefont {Pan}}, \bibinfo {author} {\bibfnamefont {Y.}~\bibnamefont {Zhang}}, \bibinfo {author} {\bibfnamefont {J.}~\bibnamefont {Kang}},\ and\ \bibinfo {author} {\bibfnamefont {Z.~Y.}\ \bibnamefont {Meng}},\ }\href {https://doi.org/10.1088/0256-307x/38/7/077305} {\bibfield  {journal} {\bibinfo  {journal} {Chinese Physics Letters}\ }\textbf {\bibinfo {volume} {38}},\ \bibinfo {pages} {077305} (\bibinfo {year} {2021}{\natexlab{a}})}\BibitemShut {NoStop}%
\bibitem [{\citenamefont {Parker}\ \emph {et~al.}(2021)\citenamefont {Parker}, \citenamefont {Soejima}, \citenamefont {Hauschild}, \citenamefont {Zaletel},\ and\ \citenamefont {Bultinck}}]{bultinck-tbg-strain-prl21}%
  \BibitemOpen
  \bibfield  {author} {\bibinfo {author} {\bibfnamefont {D.~E.}\ \bibnamefont {Parker}}, \bibinfo {author} {\bibfnamefont {T.}~\bibnamefont {Soejima}}, \bibinfo {author} {\bibfnamefont {J.}~\bibnamefont {Hauschild}}, \bibinfo {author} {\bibfnamefont {M.~P.}\ \bibnamefont {Zaletel}},\ and\ \bibinfo {author} {\bibfnamefont {N.}~\bibnamefont {Bultinck}},\ }\href {https://doi.org/10.1103/PhysRevLett.127.027601} {\bibfield  {journal} {\bibinfo  {journal} {Phys. Rev. Lett.}\ }\textbf {\bibinfo {volume} {127}},\ \bibinfo {pages} {027601} (\bibinfo {year} {2021})}\BibitemShut {NoStop}%
\bibitem [{\citenamefont {Xu}\ \emph {et~al.}(2024)\citenamefont {Xu}, \citenamefont {Li}, \citenamefont {Xu}, \citenamefont {Bi},\ and\ \citenamefont {Zhang}}]{xu2024maximally}%
  \BibitemOpen
  \bibfield  {author} {\bibinfo {author} {\bibfnamefont {C.}~\bibnamefont {Xu}}, \bibinfo {author} {\bibfnamefont {J.}~\bibnamefont {Li}}, \bibinfo {author} {\bibfnamefont {Y.}~\bibnamefont {Xu}}, \bibinfo {author} {\bibfnamefont {Z.}~\bibnamefont {Bi}},\ and\ \bibinfo {author} {\bibfnamefont {Y.}~\bibnamefont {Zhang}},\ }\href@noop {} {\bibfield  {journal} {\bibinfo  {journal} {Proceedings of the National Academy of Sciences}\ }\textbf {\bibinfo {volume} {121}},\ \bibinfo {pages} {e2316749121} (\bibinfo {year} {2024})}\BibitemShut {NoStop}%
\bibitem [{\citenamefont {Mao}\ \emph {et~al.}(2024)\citenamefont {Mao}, \citenamefont {Xu}, \citenamefont {Li}, \citenamefont {Bao}, \citenamefont {Liu}, \citenamefont {Xu}, \citenamefont {Felser}, \citenamefont {Fu},\ and\ \citenamefont {Zhang}}]{mao2024transfer}%
  \BibitemOpen
  \bibfield  {author} {\bibinfo {author} {\bibfnamefont {N.}~\bibnamefont {Mao}}, \bibinfo {author} {\bibfnamefont {C.}~\bibnamefont {Xu}}, \bibinfo {author} {\bibfnamefont {J.}~\bibnamefont {Li}}, \bibinfo {author} {\bibfnamefont {T.}~\bibnamefont {Bao}}, \bibinfo {author} {\bibfnamefont {P.}~\bibnamefont {Liu}}, \bibinfo {author} {\bibfnamefont {Y.}~\bibnamefont {Xu}}, \bibinfo {author} {\bibfnamefont {C.}~\bibnamefont {Felser}}, \bibinfo {author} {\bibfnamefont {L.}~\bibnamefont {Fu}},\ and\ \bibinfo {author} {\bibfnamefont {Y.}~\bibnamefont {Zhang}},\ }\href@noop {} {\bibfield  {journal} {\bibinfo  {journal} {Communications Physics}\ }\textbf {\bibinfo {volume} {7}},\ \bibinfo {pages} {262} (\bibinfo {year} {2024})}\BibitemShut {NoStop}%
\bibitem [{\citenamefont {Zhang}\ \emph {et~al.}(2021{\natexlab{b}})\citenamefont {Zhang}, \citenamefont {Liu},\ and\ \citenamefont {Fu}}]{zhang2021electronic}%
  \BibitemOpen
  \bibfield  {author} {\bibinfo {author} {\bibfnamefont {Y.}~\bibnamefont {Zhang}}, \bibinfo {author} {\bibfnamefont {T.}~\bibnamefont {Liu}},\ and\ \bibinfo {author} {\bibfnamefont {L.}~\bibnamefont {Fu}},\ }\href@noop {} {\bibfield  {journal} {\bibinfo  {journal} {Physical Review B}\ }\textbf {\bibinfo {volume} {103}},\ \bibinfo {pages} {155142} (\bibinfo {year} {2021}{\natexlab{b}})}\BibitemShut {NoStop}%
\bibitem [{\citenamefont {Xu}\ \emph {et~al.}(2025{\natexlab{a}})\citenamefont {Xu}, \citenamefont {Mao}, \citenamefont {Zeng},\ and\ \citenamefont {Zhang}}]{PhysRevLett.134.066601}%
  \BibitemOpen
  \bibfield  {author} {\bibinfo {author} {\bibfnamefont {C.}~\bibnamefont {Xu}}, \bibinfo {author} {\bibfnamefont {N.}~\bibnamefont {Mao}}, \bibinfo {author} {\bibfnamefont {T.}~\bibnamefont {Zeng}},\ and\ \bibinfo {author} {\bibfnamefont {Y.}~\bibnamefont {Zhang}},\ }\href {https://doi.org/10.1103/PhysRevLett.134.066601} {\bibfield  {journal} {\bibinfo  {journal} {Phys. Rev. Lett.}\ }\textbf {\bibinfo {volume} {134}},\ \bibinfo {pages} {066601} (\bibinfo {year} {2025}{\natexlab{a}})}\BibitemShut {NoStop}%
\bibitem [{\citenamefont {Xu}\ \emph {et~al.}(2023)\citenamefont {Xu}, \citenamefont {Sun}, \citenamefont {Jia}, \citenamefont {Liu}, \citenamefont {Xu}, \citenamefont {Li}, \citenamefont {Gu}, \citenamefont {Watanabe}, \citenamefont {Taniguchi}, \citenamefont {Tong}, \citenamefont {Jia}, \citenamefont {Shi}, \citenamefont {Jiang}, \citenamefont {Zhang}, \citenamefont {Liu},\ and\ \citenamefont {Li}}]{PhysRevX.13.031037}%
  \BibitemOpen
  \bibfield  {author} {\bibinfo {author} {\bibfnamefont {F.}~\bibnamefont {Xu}}, \bibinfo {author} {\bibfnamefont {Z.}~\bibnamefont {Sun}}, \bibinfo {author} {\bibfnamefont {T.}~\bibnamefont {Jia}}, \bibinfo {author} {\bibfnamefont {C.}~\bibnamefont {Liu}}, \bibinfo {author} {\bibfnamefont {C.}~\bibnamefont {Xu}}, \bibinfo {author} {\bibfnamefont {C.}~\bibnamefont {Li}}, \bibinfo {author} {\bibfnamefont {Y.}~\bibnamefont {Gu}}, \bibinfo {author} {\bibfnamefont {K.}~\bibnamefont {Watanabe}}, \bibinfo {author} {\bibfnamefont {T.}~\bibnamefont {Taniguchi}}, \bibinfo {author} {\bibfnamefont {B.}~\bibnamefont {Tong}}, \bibinfo {author} {\bibfnamefont {J.}~\bibnamefont {Jia}}, \bibinfo {author} {\bibfnamefont {Z.}~\bibnamefont {Shi}}, \bibinfo {author} {\bibfnamefont {S.}~\bibnamefont {Jiang}}, \bibinfo {author} {\bibfnamefont {Y.}~\bibnamefont {Zhang}}, \bibinfo {author} {\bibfnamefont {X.}~\bibnamefont {Liu}},\ and\ \bibinfo {author} {\bibfnamefont {T.}~\bibnamefont {Li}},\ }\href
  {https://doi.org/10.1103/PhysRevX.13.031037} {\bibfield  {journal} {\bibinfo  {journal} {Phys. Rev. X}\ }\textbf {\bibinfo {volume} {13}},\ \bibinfo {pages} {031037} (\bibinfo {year} {2023})}\BibitemShut {NoStop}%
\bibitem [{\citenamefont {Fan}\ \emph {et~al.}(2024)\citenamefont {Fan}, \citenamefont {Xiao},\ and\ \citenamefont {Yao}}]{fan2024orbital}%
  \BibitemOpen
  \bibfield  {author} {\bibinfo {author} {\bibfnamefont {F.-R.}\ \bibnamefont {Fan}}, \bibinfo {author} {\bibfnamefont {C.}~\bibnamefont {Xiao}},\ and\ \bibinfo {author} {\bibfnamefont {W.}~\bibnamefont {Yao}},\ }\href@noop {} {\bibfield  {journal} {\bibinfo  {journal} {Physical Review B}\ }\textbf {\bibinfo {volume} {109}},\ \bibinfo {pages} {L041403} (\bibinfo {year} {2024})}\BibitemShut {NoStop}%
\bibitem [{\citenamefont {Tong}\ \emph {et~al.}(2017)\citenamefont {Tong}, \citenamefont {Yu}, \citenamefont {Zhu}, \citenamefont {Wang}, \citenamefont {Xu},\ and\ \citenamefont {Yao}}]{tong2017topological}%
  \BibitemOpen
  \bibfield  {author} {\bibinfo {author} {\bibfnamefont {Q.}~\bibnamefont {Tong}}, \bibinfo {author} {\bibfnamefont {H.}~\bibnamefont {Yu}}, \bibinfo {author} {\bibfnamefont {Q.}~\bibnamefont {Zhu}}, \bibinfo {author} {\bibfnamefont {Y.}~\bibnamefont {Wang}}, \bibinfo {author} {\bibfnamefont {X.}~\bibnamefont {Xu}},\ and\ \bibinfo {author} {\bibfnamefont {W.}~\bibnamefont {Yao}},\ }\href@noop {} {\bibfield  {journal} {\bibinfo  {journal} {Nature Physics}\ }\textbf {\bibinfo {volume} {13}},\ \bibinfo {pages} {356} (\bibinfo {year} {2017})}\BibitemShut {NoStop}%
\bibitem [{\citenamefont {Reddy}\ \emph {et~al.}(2023{\natexlab{a}})\citenamefont {Reddy}, \citenamefont {Devakul},\ and\ \citenamefont {Fu}}]{PhysRevLett.131.246501}%
  \BibitemOpen
  \bibfield  {author} {\bibinfo {author} {\bibfnamefont {A.~P.}\ \bibnamefont {Reddy}}, \bibinfo {author} {\bibfnamefont {T.}~\bibnamefont {Devakul}},\ and\ \bibinfo {author} {\bibfnamefont {L.}~\bibnamefont {Fu}},\ }\href {https://doi.org/10.1103/PhysRevLett.131.246501} {\bibfield  {journal} {\bibinfo  {journal} {Phys. Rev. Lett.}\ }\textbf {\bibinfo {volume} {131}},\ \bibinfo {pages} {246501} (\bibinfo {year} {2023}{\natexlab{a}})}\BibitemShut {NoStop}%
\bibitem [{\citenamefont {Sheng}\ \emph {et~al.}(2024)\citenamefont {Sheng}, \citenamefont {Reddy}, \citenamefont {Abouelkomsan}, \citenamefont {Bergholtz},\ and\ \citenamefont {Fu}}]{PhysRevLett.133.066601}%
  \BibitemOpen
  \bibfield  {author} {\bibinfo {author} {\bibfnamefont {D.~N.}\ \bibnamefont {Sheng}}, \bibinfo {author} {\bibfnamefont {A.~P.}\ \bibnamefont {Reddy}}, \bibinfo {author} {\bibfnamefont {A.}~\bibnamefont {Abouelkomsan}}, \bibinfo {author} {\bibfnamefont {E.~J.}\ \bibnamefont {Bergholtz}},\ and\ \bibinfo {author} {\bibfnamefont {L.}~\bibnamefont {Fu}},\ }\href {https://doi.org/10.1103/PhysRevLett.133.066601} {\bibfield  {journal} {\bibinfo  {journal} {Phys. Rev. Lett.}\ }\textbf {\bibinfo {volume} {133}},\ \bibinfo {pages} {066601} (\bibinfo {year} {2024})}\BibitemShut {NoStop}%
\bibitem [{\citenamefont {Xu}\ \emph {et~al.}(2025{\natexlab{b}})\citenamefont {Xu}, \citenamefont {Chang}, \citenamefont {Xiao}, \citenamefont {Zhang}, \citenamefont {Liu}, \citenamefont {Sun}, \citenamefont {Mao}, \citenamefont {Peshcherenko}, \citenamefont {Li}, \citenamefont {Watanabe} \emph {et~al.}}]{xu2025interplay}%
  \BibitemOpen
  \bibfield  {author} {\bibinfo {author} {\bibfnamefont {F.}~\bibnamefont {Xu}}, \bibinfo {author} {\bibfnamefont {X.}~\bibnamefont {Chang}}, \bibinfo {author} {\bibfnamefont {J.}~\bibnamefont {Xiao}}, \bibinfo {author} {\bibfnamefont {Y.}~\bibnamefont {Zhang}}, \bibinfo {author} {\bibfnamefont {F.}~\bibnamefont {Liu}}, \bibinfo {author} {\bibfnamefont {Z.}~\bibnamefont {Sun}}, \bibinfo {author} {\bibfnamefont {N.}~\bibnamefont {Mao}}, \bibinfo {author} {\bibfnamefont {N.}~\bibnamefont {Peshcherenko}}, \bibinfo {author} {\bibfnamefont {J.}~\bibnamefont {Li}}, \bibinfo {author} {\bibfnamefont {K.}~\bibnamefont {Watanabe}}, \emph {et~al.},\ }\href@noop {} {\bibfield  {journal} {\bibinfo  {journal} {Nature Physics}\ ,\ \bibinfo {pages} {1}} (\bibinfo {year} {2025}{\natexlab{b}})}\BibitemShut {NoStop}%
\bibitem [{\citenamefont {Zhang}\ \emph {et~al.}(2024)\citenamefont {Zhang}, \citenamefont {Wang}, \citenamefont {Liu}, \citenamefont {Fan}, \citenamefont {Cao},\ and\ \citenamefont {Xiao}}]{zhang2024polarization}%
  \BibitemOpen
  \bibfield  {author} {\bibinfo {author} {\bibfnamefont {X.-W.}\ \bibnamefont {Zhang}}, \bibinfo {author} {\bibfnamefont {C.}~\bibnamefont {Wang}}, \bibinfo {author} {\bibfnamefont {X.}~\bibnamefont {Liu}}, \bibinfo {author} {\bibfnamefont {Y.}~\bibnamefont {Fan}}, \bibinfo {author} {\bibfnamefont {T.}~\bibnamefont {Cao}},\ and\ \bibinfo {author} {\bibfnamefont {D.}~\bibnamefont {Xiao}},\ }\href@noop {} {\bibfield  {journal} {\bibinfo  {journal} {Nature Communications}\ }\textbf {\bibinfo {volume} {15}},\ \bibinfo {pages} {4223} (\bibinfo {year} {2024})}\BibitemShut {NoStop}%
\bibitem [{\citenamefont {Reddy}\ \emph {et~al.}(2023{\natexlab{b}})\citenamefont {Reddy}, \citenamefont {Alsallom}, \citenamefont {Zhang}, \citenamefont {Devakul},\ and\ \citenamefont {Fu}}]{PhysRevB.108.085117}%
  \BibitemOpen
  \bibfield  {author} {\bibinfo {author} {\bibfnamefont {A.~P.}\ \bibnamefont {Reddy}}, \bibinfo {author} {\bibfnamefont {F.}~\bibnamefont {Alsallom}}, \bibinfo {author} {\bibfnamefont {Y.}~\bibnamefont {Zhang}}, \bibinfo {author} {\bibfnamefont {T.}~\bibnamefont {Devakul}},\ and\ \bibinfo {author} {\bibfnamefont {L.}~\bibnamefont {Fu}},\ }\href {https://doi.org/10.1103/PhysRevB.108.085117} {\bibfield  {journal} {\bibinfo  {journal} {Phys. Rev. B}\ }\textbf {\bibinfo {volume} {108}},\ \bibinfo {pages} {085117} (\bibinfo {year} {2023}{\natexlab{b}})}\BibitemShut {NoStop}%
\bibitem [{\citenamefont {Xie}\ \emph {et~al.}(2022)\citenamefont {Xie}, \citenamefont {Zhang}, \citenamefont {Hu}, \citenamefont {Mak},\ and\ \citenamefont {Law}}]{PhysRevLett.128.026402}%
  \BibitemOpen
  \bibfield  {author} {\bibinfo {author} {\bibfnamefont {Y.-M.}\ \bibnamefont {Xie}}, \bibinfo {author} {\bibfnamefont {C.-P.}\ \bibnamefont {Zhang}}, \bibinfo {author} {\bibfnamefont {J.-X.}\ \bibnamefont {Hu}}, \bibinfo {author} {\bibfnamefont {K.~F.}\ \bibnamefont {Mak}},\ and\ \bibinfo {author} {\bibfnamefont {K.~T.}\ \bibnamefont {Law}},\ }\href {https://doi.org/10.1103/PhysRevLett.128.026402} {\bibfield  {journal} {\bibinfo  {journal} {Phys. Rev. Lett.}\ }\textbf {\bibinfo {volume} {128}},\ \bibinfo {pages} {026402} (\bibinfo {year} {2022})}\BibitemShut {NoStop}%
\bibitem [{\citenamefont {Pan}\ \emph {et~al.}(2022)\citenamefont {Pan}, \citenamefont {Xie}, \citenamefont {Wu},\ and\ \citenamefont {Das~Sarma}}]{PhysRevLett.129.056804}%
  \BibitemOpen
  \bibfield  {author} {\bibinfo {author} {\bibfnamefont {H.}~\bibnamefont {Pan}}, \bibinfo {author} {\bibfnamefont {M.}~\bibnamefont {Xie}}, \bibinfo {author} {\bibfnamefont {F.}~\bibnamefont {Wu}},\ and\ \bibinfo {author} {\bibfnamefont {S.}~\bibnamefont {Das~Sarma}},\ }\href {https://doi.org/10.1103/PhysRevLett.129.056804} {\bibfield  {journal} {\bibinfo  {journal} {Phys. Rev. Lett.}\ }\textbf {\bibinfo {volume} {129}},\ \bibinfo {pages} {056804} (\bibinfo {year} {2022})}\BibitemShut {NoStop}%
\bibitem [{\citenamefont {Li}\ and\ \citenamefont {Wu}(2025)}]{PhysRevB.111.125122}%
  \BibitemOpen
  \bibfield  {author} {\bibinfo {author} {\bibfnamefont {B.}~\bibnamefont {Li}}\ and\ \bibinfo {author} {\bibfnamefont {F.}~\bibnamefont {Wu}},\ }\href {https://doi.org/10.1103/PhysRevB.111.125122} {\bibfield  {journal} {\bibinfo  {journal} {Phys. Rev. B}\ }\textbf {\bibinfo {volume} {111}},\ \bibinfo {pages} {125122} (\bibinfo {year} {2025})}\BibitemShut {NoStop}%
\bibitem [{\citenamefont {Qiu}\ \emph {et~al.}(2023)\citenamefont {Qiu}, \citenamefont {Li}, \citenamefont {Luo},\ and\ \citenamefont {Wu}}]{PhysRevX.13.041026}%
  \BibitemOpen
  \bibfield  {author} {\bibinfo {author} {\bibfnamefont {W.-X.}\ \bibnamefont {Qiu}}, \bibinfo {author} {\bibfnamefont {B.}~\bibnamefont {Li}}, \bibinfo {author} {\bibfnamefont {X.-J.}\ \bibnamefont {Luo}},\ and\ \bibinfo {author} {\bibfnamefont {F.}~\bibnamefont {Wu}},\ }\href {https://doi.org/10.1103/PhysRevX.13.041026} {\bibfield  {journal} {\bibinfo  {journal} {Phys. Rev. X}\ }\textbf {\bibinfo {volume} {13}},\ \bibinfo {pages} {041026} (\bibinfo {year} {2023})}\BibitemShut {NoStop}%
\bibitem [{\citenamefont {Wu}\ \emph {et~al.}(2019)\citenamefont {Wu}, \citenamefont {Lovorn}, \citenamefont {Tutuc}, \citenamefont {Martin},\ and\ \citenamefont {MacDonald}}]{PhysRevLett.122.086402}%
  \BibitemOpen
  \bibfield  {author} {\bibinfo {author} {\bibfnamefont {F.}~\bibnamefont {Wu}}, \bibinfo {author} {\bibfnamefont {T.}~\bibnamefont {Lovorn}}, \bibinfo {author} {\bibfnamefont {E.}~\bibnamefont {Tutuc}}, \bibinfo {author} {\bibfnamefont {I.}~\bibnamefont {Martin}},\ and\ \bibinfo {author} {\bibfnamefont {A.~H.}\ \bibnamefont {MacDonald}},\ }\href {https://doi.org/10.1103/PhysRevLett.122.086402} {\bibfield  {journal} {\bibinfo  {journal} {Phys. Rev. Lett.}\ }\textbf {\bibinfo {volume} {122}},\ \bibinfo {pages} {086402} (\bibinfo {year} {2019})}\BibitemShut {NoStop}%
\bibitem [{\citenamefont {Li}\ \emph {et~al.}(2024)\citenamefont {Li}, \citenamefont {Qiu},\ and\ \citenamefont {Wu}}]{PhysRevB.109.L041106}%
  \BibitemOpen
  \bibfield  {author} {\bibinfo {author} {\bibfnamefont {B.}~\bibnamefont {Li}}, \bibinfo {author} {\bibfnamefont {W.-X.}\ \bibnamefont {Qiu}},\ and\ \bibinfo {author} {\bibfnamefont {F.}~\bibnamefont {Wu}},\ }\href {https://doi.org/10.1103/PhysRevB.109.L041106} {\bibfield  {journal} {\bibinfo  {journal} {Phys. Rev. B}\ }\textbf {\bibinfo {volume} {109}},\ \bibinfo {pages} {L041106} (\bibinfo {year} {2024})}\BibitemShut {NoStop}%
\bibitem [{\citenamefont {Cai}\ \emph {et~al.}(2023)\citenamefont {Cai}, \citenamefont {Anderson}, \citenamefont {Wang}, \citenamefont {Zhang}, \citenamefont {Liu}, \citenamefont {Holtzmann}, \citenamefont {Zhang}, \citenamefont {Fan}, \citenamefont {Taniguchi}, \citenamefont {Watanabe} \emph {et~al.}}]{cai2023signatures}%
  \BibitemOpen
  \bibfield  {author} {\bibinfo {author} {\bibfnamefont {J.}~\bibnamefont {Cai}}, \bibinfo {author} {\bibfnamefont {E.}~\bibnamefont {Anderson}}, \bibinfo {author} {\bibfnamefont {C.}~\bibnamefont {Wang}}, \bibinfo {author} {\bibfnamefont {X.}~\bibnamefont {Zhang}}, \bibinfo {author} {\bibfnamefont {X.}~\bibnamefont {Liu}}, \bibinfo {author} {\bibfnamefont {W.}~\bibnamefont {Holtzmann}}, \bibinfo {author} {\bibfnamefont {Y.}~\bibnamefont {Zhang}}, \bibinfo {author} {\bibfnamefont {F.}~\bibnamefont {Fan}}, \bibinfo {author} {\bibfnamefont {T.}~\bibnamefont {Taniguchi}}, \bibinfo {author} {\bibfnamefont {K.}~\bibnamefont {Watanabe}}, \emph {et~al.},\ }\href@noop {} {\bibfield  {journal} {\bibinfo  {journal} {Nature}\ }\textbf {\bibinfo {volume} {622}},\ \bibinfo {pages} {63} (\bibinfo {year} {2023})}\BibitemShut {NoStop}%
\bibitem [{\citenamefont {Zhang}\ \emph {et~al.}(2014)\citenamefont {Zhang}, \citenamefont {Liu}, \citenamefont {Wong}, \citenamefont {Kim}, \citenamefont {Hong}, \citenamefont {Liu}, \citenamefont {Cao}, \citenamefont {Louie}, \citenamefont {Wang},\ and\ \citenamefont {Yang}}]{zhang2014three}%
  \BibitemOpen
  \bibfield  {author} {\bibinfo {author} {\bibfnamefont {L.}~\bibnamefont {Zhang}}, \bibinfo {author} {\bibfnamefont {K.}~\bibnamefont {Liu}}, \bibinfo {author} {\bibfnamefont {A.~B.}\ \bibnamefont {Wong}}, \bibinfo {author} {\bibfnamefont {J.}~\bibnamefont {Kim}}, \bibinfo {author} {\bibfnamefont {X.}~\bibnamefont {Hong}}, \bibinfo {author} {\bibfnamefont {C.}~\bibnamefont {Liu}}, \bibinfo {author} {\bibfnamefont {T.}~\bibnamefont {Cao}}, \bibinfo {author} {\bibfnamefont {S.~G.}\ \bibnamefont {Louie}}, \bibinfo {author} {\bibfnamefont {F.}~\bibnamefont {Wang}},\ and\ \bibinfo {author} {\bibfnamefont {P.}~\bibnamefont {Yang}},\ }\href@noop {} {\bibfield  {journal} {\bibinfo  {journal} {Nano letters}\ }\textbf {\bibinfo {volume} {14}},\ \bibinfo {pages} {6418} (\bibinfo {year} {2014})}\BibitemShut {NoStop}%
\bibitem [{\citenamefont {Ci}\ \emph {et~al.}(2022)\citenamefont {Ci}, \citenamefont {Zhao}, \citenamefont {Sun}, \citenamefont {Rho}, \citenamefont {Chen}, \citenamefont {Grigoropoulos}, \citenamefont {Jin}, \citenamefont {Li},\ and\ \citenamefont {Wu}}]{ci2022breaking}%
  \BibitemOpen
  \bibfield  {author} {\bibinfo {author} {\bibfnamefont {P.}~\bibnamefont {Ci}}, \bibinfo {author} {\bibfnamefont {Y.}~\bibnamefont {Zhao}}, \bibinfo {author} {\bibfnamefont {M.}~\bibnamefont {Sun}}, \bibinfo {author} {\bibfnamefont {Y.}~\bibnamefont {Rho}}, \bibinfo {author} {\bibfnamefont {Y.}~\bibnamefont {Chen}}, \bibinfo {author} {\bibfnamefont {C.~P.}\ \bibnamefont {Grigoropoulos}}, \bibinfo {author} {\bibfnamefont {S.}~\bibnamefont {Jin}}, \bibinfo {author} {\bibfnamefont {X.}~\bibnamefont {Li}},\ and\ \bibinfo {author} {\bibfnamefont {J.}~\bibnamefont {Wu}},\ }\href@noop {} {\bibfield  {journal} {\bibinfo  {journal} {Nano letters}\ }\textbf {\bibinfo {volume} {22}},\ \bibinfo {pages} {9027} (\bibinfo {year} {2022})}\BibitemShut {NoStop}%
\bibitem [{\citenamefont {Fan}\ \emph {et~al.}(2017)\citenamefont {Fan}, \citenamefont {Jiang}, \citenamefont {Zhuang}, \citenamefont {Liu}, \citenamefont {Xu}, \citenamefont {Zheng}, \citenamefont {Fan}, \citenamefont {Li}, \citenamefont {Wu}, \citenamefont {Zhu} \emph {et~al.}}]{fan2017broken}%
  \BibitemOpen
  \bibfield  {author} {\bibinfo {author} {\bibfnamefont {X.}~\bibnamefont {Fan}}, \bibinfo {author} {\bibfnamefont {Y.}~\bibnamefont {Jiang}}, \bibinfo {author} {\bibfnamefont {X.}~\bibnamefont {Zhuang}}, \bibinfo {author} {\bibfnamefont {H.}~\bibnamefont {Liu}}, \bibinfo {author} {\bibfnamefont {T.}~\bibnamefont {Xu}}, \bibinfo {author} {\bibfnamefont {W.}~\bibnamefont {Zheng}}, \bibinfo {author} {\bibfnamefont {P.}~\bibnamefont {Fan}}, \bibinfo {author} {\bibfnamefont {H.}~\bibnamefont {Li}}, \bibinfo {author} {\bibfnamefont {X.}~\bibnamefont {Wu}}, \bibinfo {author} {\bibfnamefont {X.}~\bibnamefont {Zhu}}, \emph {et~al.},\ }\href@noop {} {\bibfield  {journal} {\bibinfo  {journal} {ACS nano}\ }\textbf {\bibinfo {volume} {11}},\ \bibinfo {pages} {4892} (\bibinfo {year} {2017})}\BibitemShut {NoStop}%
\bibitem [{\citenamefont {Chen}\ \emph {et~al.}(2025)\citenamefont {Chen}, \citenamefont {Bai}, \citenamefont {Qi}, \citenamefont {Zhang}, \citenamefont {Qin}, \citenamefont {Fan},\ and\ \citenamefont {Xiao}}]{chen2025structure}%
  \BibitemOpen
  \bibfield  {author} {\bibinfo {author} {\bibfnamefont {J.}~\bibnamefont {Chen}}, \bibinfo {author} {\bibfnamefont {Y.}~\bibnamefont {Bai}}, \bibinfo {author} {\bibfnamefont {M.}~\bibnamefont {Qi}}, \bibinfo {author} {\bibfnamefont {W.}~\bibnamefont {Zhang}}, \bibinfo {author} {\bibfnamefont {C.}~\bibnamefont {Qin}}, \bibinfo {author} {\bibfnamefont {X.}~\bibnamefont {Fan}},\ and\ \bibinfo {author} {\bibfnamefont {L.}~\bibnamefont {Xiao}},\ }\href@noop {} {\bibfield  {journal} {\bibinfo  {journal} {Advanced Materials}\ }\textbf {\bibinfo {volume} {37}},\ \bibinfo {pages} {2415214} (\bibinfo {year} {2025})}\BibitemShut {NoStop}%
\bibitem [{\citenamefont {Lu}\ \emph {et~al.}(2024)\citenamefont {Lu}, \citenamefont {Xie}, \citenamefont {Yang}, \citenamefont {Zhang}, \citenamefont {Kong}, \citenamefont {Li}, \citenamefont {Ding}, \citenamefont {Wang},\ and\ \citenamefont {Liu}}]{lu2024magic}%
  \BibitemOpen
  \bibfield  {author} {\bibinfo {author} {\bibfnamefont {X.}~\bibnamefont {Lu}}, \bibinfo {author} {\bibfnamefont {B.}~\bibnamefont {Xie}}, \bibinfo {author} {\bibfnamefont {Y.}~\bibnamefont {Yang}}, \bibinfo {author} {\bibfnamefont {Y.}~\bibnamefont {Zhang}}, \bibinfo {author} {\bibfnamefont {X.}~\bibnamefont {Kong}}, \bibinfo {author} {\bibfnamefont {J.}~\bibnamefont {Li}}, \bibinfo {author} {\bibfnamefont {F.}~\bibnamefont {Ding}}, \bibinfo {author} {\bibfnamefont {Z.-J.}\ \bibnamefont {Wang}},\ and\ \bibinfo {author} {\bibfnamefont {J.}~\bibnamefont {Liu}},\ }\href@noop {} {\bibfield  {journal} {\bibinfo  {journal} {Physical Review Letters}\ }\textbf {\bibinfo {volume} {132}},\ \bibinfo {pages} {056601} (\bibinfo {year} {2024})}\BibitemShut {NoStop}%
\bibitem [{\citenamefont {Staley}\ \emph {et~al.}(2009)\citenamefont {Staley}, \citenamefont {Wu}, \citenamefont {Eklund}, \citenamefont {Liu}, \citenamefont {Li},\ and\ \citenamefont {Xu}}]{NbSe2_SC}%
  \BibitemOpen
  \bibfield  {author} {\bibinfo {author} {\bibfnamefont {N.~E.}\ \bibnamefont {Staley}}, \bibinfo {author} {\bibfnamefont {J.}~\bibnamefont {Wu}}, \bibinfo {author} {\bibfnamefont {P.}~\bibnamefont {Eklund}}, \bibinfo {author} {\bibfnamefont {Y.}~\bibnamefont {Liu}}, \bibinfo {author} {\bibfnamefont {L.}~\bibnamefont {Li}},\ and\ \bibinfo {author} {\bibfnamefont {Z.}~\bibnamefont {Xu}},\ }\href {https://doi.org/10.1103/PhysRevB.80.184505} {\bibfield  {journal} {\bibinfo  {journal} {Phys. Rev. B}\ }\textbf {\bibinfo {volume} {80}},\ \bibinfo {pages} {184505} (\bibinfo {year} {2009})}\BibitemShut {NoStop}%
\bibitem [{\citenamefont {Ozaki}(2003)}]{OpenMX1}%
  \BibitemOpen
  \bibfield  {author} {\bibinfo {author} {\bibfnamefont {T.}~\bibnamefont {Ozaki}},\ }\href {https://doi.org/10.1103/PhysRevB.67.155108} {\bibfield  {journal} {\bibinfo  {journal} {Phys. Rev. B}\ }\textbf {\bibinfo {volume} {67}},\ \bibinfo {pages} {155108} (\bibinfo {year} {2003})}\BibitemShut {NoStop}%
\bibitem [{\citenamefont {Ozaki}\ and\ \citenamefont {Kino}(2004)}]{OpenMX2}%
  \BibitemOpen
  \bibfield  {author} {\bibinfo {author} {\bibfnamefont {T.}~\bibnamefont {Ozaki}}\ and\ \bibinfo {author} {\bibfnamefont {H.}~\bibnamefont {Kino}},\ }\href {https://doi.org/10.1103/PhysRevB.69.195113} {\bibfield  {journal} {\bibinfo  {journal} {Phys. Rev. B}\ }\textbf {\bibinfo {volume} {69}},\ \bibinfo {pages} {195113} (\bibinfo {year} {2004})}\BibitemShut {NoStop}%
\bibitem [{\citenamefont {Grimme}\ \emph {et~al.}(2010)\citenamefont {Grimme}, \citenamefont {Antony}, \citenamefont {Ehrlich},\ and\ \citenamefont {Krieg}}]{OpenMX3}%
  \BibitemOpen
  \bibfield  {author} {\bibinfo {author} {\bibfnamefont {S.}~\bibnamefont {Grimme}}, \bibinfo {author} {\bibfnamefont {J.}~\bibnamefont {Antony}}, \bibinfo {author} {\bibfnamefont {S.}~\bibnamefont {Ehrlich}},\ and\ \bibinfo {author} {\bibfnamefont {H.}~\bibnamefont {Krieg}},\ }\href {https://doi.org/10.1063/1.3382344} {\bibfield  {journal} {\bibinfo  {journal} {The Journal of Chemical Physics}\ }\textbf {\bibinfo {volume} {132}},\ \bibinfo {pages} {154104} (\bibinfo {year} {2010})}\BibitemShut {NoStop}%
\bibitem [{\citenamefont {Michaud-Rioux}\ \emph {et~al.}(2016)\citenamefont {Michaud-Rioux}, \citenamefont {Zhang},\ and\ \citenamefont {Guo}}]{RESCU}%
  \BibitemOpen
  \bibfield  {author} {\bibinfo {author} {\bibfnamefont {V.}~\bibnamefont {Michaud-Rioux}}, \bibinfo {author} {\bibfnamefont {L.}~\bibnamefont {Zhang}},\ and\ \bibinfo {author} {\bibfnamefont {H.}~\bibnamefont {Guo}},\ }\href {https://doi.org/https://doi.org/10.1016/j.jcp.2015.12.014} {\bibfield  {journal} {\bibinfo  {journal} {Journal of Computational Physics}\ }\textbf {\bibinfo {volume} {307}},\ \bibinfo {pages} {593} (\bibinfo {year} {2016})}\BibitemShut {NoStop}%
\bibitem [{\citenamefont {Kresse}\ and\ \citenamefont {Hafner}(1993)}]{vasp_Kresse1993}%
  \BibitemOpen
  \bibfield  {author} {\bibinfo {author} {\bibfnamefont {G.}~\bibnamefont {Kresse}}\ and\ \bibinfo {author} {\bibfnamefont {J.}~\bibnamefont {Hafner}},\ }\href {https://doi.org/10.1103/physrevb.47.558} {\bibfield  {journal} {\bibinfo  {journal} {Physical Review B}\ }\textbf {\bibinfo {volume} {47}},\ \bibinfo {pages} {558–561} (\bibinfo {year} {1993})}\BibitemShut {NoStop}%
\bibitem [{\citenamefont {Kresse}\ and\ \citenamefont {Hafner}(1994)}]{vasp_Kresse1994}%
  \BibitemOpen
  \bibfield  {author} {\bibinfo {author} {\bibfnamefont {G.}~\bibnamefont {Kresse}}\ and\ \bibinfo {author} {\bibfnamefont {J.}~\bibnamefont {Hafner}},\ }\href {https://doi.org/10.1103/physrevb.49.14251} {\bibfield  {journal} {\bibinfo  {journal} {Physical Review B}\ }\textbf {\bibinfo {volume} {49}},\ \bibinfo {pages} {14251–14269} (\bibinfo {year} {1994})}\BibitemShut {NoStop}%
\bibitem [{\citenamefont {Kresse}\ and\ \citenamefont {Furthm\"{u}ller}(1996)}]{vasp_Kresse1996}%
  \BibitemOpen
  \bibfield  {author} {\bibinfo {author} {\bibfnamefont {G.}~\bibnamefont {Kresse}}\ and\ \bibinfo {author} {\bibfnamefont {J.}~\bibnamefont {Furthm\"{u}ller}},\ }\href {https://doi.org/10.1103/physrevb.54.11169} {\bibfield  {journal} {\bibinfo  {journal} {Physical Review B}\ }\textbf {\bibinfo {volume} {54}},\ \bibinfo {pages} {11169–11186} (\bibinfo {year} {1996})}\BibitemShut {NoStop}%
\bibitem [{\citenamefont {Kresse}\ and\ \citenamefont {Joubert}(1999)}]{vasp_Kresse1999}%
  \BibitemOpen
  \bibfield  {author} {\bibinfo {author} {\bibfnamefont {G.}~\bibnamefont {Kresse}}\ and\ \bibinfo {author} {\bibfnamefont {D.}~\bibnamefont {Joubert}},\ }\href {https://doi.org/10.1103/physrevb.59.1758} {\bibfield  {journal} {\bibinfo  {journal} {Physical Review B}\ }\textbf {\bibinfo {volume} {59}},\ \bibinfo {pages} {1758–1775} (\bibinfo {year} {1999})}\BibitemShut {NoStop}%
\bibitem [{\citenamefont {Moon}\ and\ \citenamefont {Koshino}(2012)}]{twisted1}%
  \BibitemOpen
  \bibfield  {author} {\bibinfo {author} {\bibfnamefont {P.}~\bibnamefont {Moon}}\ and\ \bibinfo {author} {\bibfnamefont {M.}~\bibnamefont {Koshino}},\ }\href {https://doi.org/10.1103/PhysRevB.85.195458} {\bibfield  {journal} {\bibinfo  {journal} {Phys. Rev. B}\ }\textbf {\bibinfo {volume} {85}},\ \bibinfo {pages} {195458} (\bibinfo {year} {2012})}\BibitemShut {NoStop}%
\bibitem [{\citenamefont {Mele}(2010)}]{twisted2}%
  \BibitemOpen
  \bibfield  {author} {\bibinfo {author} {\bibfnamefont {E.~J.}\ \bibnamefont {Mele}},\ }\href {https://doi.org/10.1103/PhysRevB.81.161405} {\bibfield  {journal} {\bibinfo  {journal} {Phys. Rev. B}\ }\textbf {\bibinfo {volume} {81}},\ \bibinfo {pages} {161405} (\bibinfo {year} {2010})}\BibitemShut {NoStop}%
\bibitem [{\citenamefont {Cheung}\ \emph {et~al.}(2024)\citenamefont {Cheung}, \citenamefont {Goodwin}, \citenamefont {Han}, \citenamefont {Lu}, \citenamefont {Mostofi},\ and\ \citenamefont {Lischner}}]{vortexCDW}%
  \BibitemOpen
  \bibfield  {author} {\bibinfo {author} {\bibfnamefont {C.~T.~S.}\ \bibnamefont {Cheung}}, \bibinfo {author} {\bibfnamefont {Z.~A.~H.}\ \bibnamefont {Goodwin}}, \bibinfo {author} {\bibfnamefont {Y.}~\bibnamefont {Han}}, \bibinfo {author} {\bibfnamefont {J.}~\bibnamefont {Lu}}, \bibinfo {author} {\bibfnamefont {A.~A.}\ \bibnamefont {Mostofi}},\ and\ \bibinfo {author} {\bibfnamefont {J.}~\bibnamefont {Lischner}},\ }\href {https://doi.org/10.1021/acs.nanolett.4c02750} {\bibfield  {journal} {\bibinfo  {journal} {Nano Letters}\ }\textbf {\bibinfo {volume} {24}},\ \bibinfo {pages} {12088} (\bibinfo {year} {2024})},\ \bibinfo {note} {pMID: 39297477}\BibitemShut {NoStop}%
\bibitem [{\citenamefont {Guinea}\ and\ \citenamefont {Walet}(2019)}]{TBG_gap_Guinea2019}%
  \BibitemOpen
  \bibfield  {author} {\bibinfo {author} {\bibfnamefont {F.}~\bibnamefont {Guinea}}\ and\ \bibinfo {author} {\bibfnamefont {N.~R.}\ \bibnamefont {Walet}},\ }\bibfield  {journal} {\bibinfo  {journal} {Physical Review B}\ }\textbf {\bibinfo {volume} {99}},\ \href {https://doi.org/10.1103/physrevb.99.205134} {10.1103/physrevb.99.205134} (\bibinfo {year} {2019})\BibitemShut {NoStop}%
\bibitem [{\citenamefont {Miao}\ \emph {et~al.}(2023)\citenamefont {Miao}, \citenamefont {Li}, \citenamefont {Han}, \citenamefont {Pan},\ and\ \citenamefont {Dai}}]{TBG_gap_Miao2023}%
  \BibitemOpen
  \bibfield  {author} {\bibinfo {author} {\bibfnamefont {W.}~\bibnamefont {Miao}}, \bibinfo {author} {\bibfnamefont {C.}~\bibnamefont {Li}}, \bibinfo {author} {\bibfnamefont {X.}~\bibnamefont {Han}}, \bibinfo {author} {\bibfnamefont {D.}~\bibnamefont {Pan}},\ and\ \bibinfo {author} {\bibfnamefont {X.}~\bibnamefont {Dai}},\ }\bibfield  {journal} {\bibinfo  {journal} {Physical Review B}\ }\textbf {\bibinfo {volume} {107}},\ \href {https://doi.org/10.1103/physrevb.107.125112} {10.1103/physrevb.107.125112} (\bibinfo {year} {2023})\BibitemShut {NoStop}%
\bibitem [{\citenamefont {Uchida}\ \emph {et~al.}(2014)\citenamefont {Uchida}, \citenamefont {Furuya}, \citenamefont {Iwata},\ and\ \citenamefont {Oshiyama}}]{TBG_gap_Uchida2014}%
  \BibitemOpen
  \bibfield  {author} {\bibinfo {author} {\bibfnamefont {K.}~\bibnamefont {Uchida}}, \bibinfo {author} {\bibfnamefont {S.}~\bibnamefont {Furuya}}, \bibinfo {author} {\bibfnamefont {J.-I.}\ \bibnamefont {Iwata}},\ and\ \bibinfo {author} {\bibfnamefont {A.}~\bibnamefont {Oshiyama}},\ }\bibfield  {journal} {\bibinfo  {journal} {Physical Review B}\ }\textbf {\bibinfo {volume} {90}},\ \href {https://doi.org/10.1103/physrevb.90.155451} {10.1103/physrevb.90.155451} (\bibinfo {year} {2014})\BibitemShut {NoStop}%
\bibitem [{\citenamefont {Xie}\ and\ \citenamefont {Liu}(2023)}]{TBG_gap_Xie2023}%
  \BibitemOpen
  \bibfield  {author} {\bibinfo {author} {\bibfnamefont {B.}~\bibnamefont {Xie}}\ and\ \bibinfo {author} {\bibfnamefont {J.}~\bibnamefont {Liu}},\ }\bibfield  {journal} {\bibinfo  {journal} {Physical Review B}\ }\textbf {\bibinfo {volume} {108}},\ \href {https://doi.org/10.1103/physrevb.108.094115} {10.1103/physrevb.108.094115} (\bibinfo {year} {2023})\BibitemShut {NoStop}%
\bibitem [{\citenamefont {Wickramaratne}\ \emph {et~al.}(2020)\citenamefont {Wickramaratne}, \citenamefont {Khmelevskyi}, \citenamefont {Agterberg},\ and\ \citenamefont {Mazin}}]{NbSe2_orbit_Wickramaratne2020}%
  \BibitemOpen
  \bibfield  {author} {\bibinfo {author} {\bibfnamefont {D.}~\bibnamefont {Wickramaratne}}, \bibinfo {author} {\bibfnamefont {S.}~\bibnamefont {Khmelevskyi}}, \bibinfo {author} {\bibfnamefont {D.~F.}\ \bibnamefont {Agterberg}},\ and\ \bibinfo {author} {\bibfnamefont {I.}~\bibnamefont {Mazin}},\ }\bibfield  {journal} {\bibinfo  {journal} {Physical Review X}\ }\textbf {\bibinfo {volume} {10}},\ \href {https://doi.org/10.1103/physrevx.10.041003} {10.1103/physrevx.10.041003} (\bibinfo {year} {2020})\BibitemShut {NoStop}%
\bibitem [{\citenamefont {Enaldiev}\ \emph {et~al.}(2020)\citenamefont {Enaldiev}, \citenamefont {Zólyomi}, \citenamefont {Yelgel}, \citenamefont {Magorrian},\ and\ \citenamefont {Fal’ko}}]{stack_MX_Enaldiev2020}%
  \BibitemOpen
  \bibfield  {author} {\bibinfo {author} {\bibfnamefont {V.}~\bibnamefont {Enaldiev}}, \bibinfo {author} {\bibfnamefont {V.}~\bibnamefont {Zólyomi}}, \bibinfo {author} {\bibfnamefont {C.}~\bibnamefont {Yelgel}}, \bibinfo {author} {\bibfnamefont {S.}~\bibnamefont {Magorrian}},\ and\ \bibinfo {author} {\bibfnamefont {V.}~\bibnamefont {Fal’ko}},\ }\bibfield  {journal} {\bibinfo  {journal} {Physical Review Letters}\ }\textbf {\bibinfo {volume} {124}},\ \href {https://doi.org/10.1103/physrevlett.124.206101} {10.1103/physrevlett.124.206101} (\bibinfo {year} {2020})\BibitemShut {NoStop}%
\bibitem [{\citenamefont {Kim}\ \emph {et~al.}(2023)\citenamefont {Kim}, \citenamefont {Choi}, \citenamefont {Lantagne-Hurtubise}, \citenamefont {Lewandowski}, \citenamefont {Thomson}, \citenamefont {Kong}, \citenamefont {Zhou}, \citenamefont {Baum}, \citenamefont {Zhang}, \citenamefont {Holleis}, \citenamefont {Watanabe}, \citenamefont {Taniguchi}, \citenamefont {Young}, \citenamefont {Alicea},\ and\ \citenamefont {Nadj-Perge}}]{Kim2023}%
  \BibitemOpen
  \bibfield  {author} {\bibinfo {author} {\bibfnamefont {H.}~\bibnamefont {Kim}}, \bibinfo {author} {\bibfnamefont {Y.}~\bibnamefont {Choi}}, \bibinfo {author} {\bibfnamefont {{\'E}.}~\bibnamefont {Lantagne-Hurtubise}}, \bibinfo {author} {\bibfnamefont {C.}~\bibnamefont {Lewandowski}}, \bibinfo {author} {\bibfnamefont {A.}~\bibnamefont {Thomson}}, \bibinfo {author} {\bibfnamefont {L.}~\bibnamefont {Kong}}, \bibinfo {author} {\bibfnamefont {H.}~\bibnamefont {Zhou}}, \bibinfo {author} {\bibfnamefont {E.}~\bibnamefont {Baum}}, \bibinfo {author} {\bibfnamefont {Y.}~\bibnamefont {Zhang}}, \bibinfo {author} {\bibfnamefont {L.}~\bibnamefont {Holleis}}, \bibinfo {author} {\bibfnamefont {K.}~\bibnamefont {Watanabe}}, \bibinfo {author} {\bibfnamefont {T.}~\bibnamefont {Taniguchi}}, \bibinfo {author} {\bibfnamefont {A.~F.}\ \bibnamefont {Young}}, \bibinfo {author} {\bibfnamefont {J.}~\bibnamefont {Alicea}},\ and\ \bibinfo {author} {\bibfnamefont {S.}~\bibnamefont {Nadj-Perge}},\ }\href
  {https://doi.org/10.1038/s41586-023-06663-8} {\bibfield  {journal} {\bibinfo  {journal} {Nature}\ }\textbf {\bibinfo {volume} {623}},\ \bibinfo {pages} {942} (\bibinfo {year} {2023})}\BibitemShut {NoStop}%
\bibitem [{\citenamefont {Shen}\ \emph {et~al.}(2023)\citenamefont {Shen}, \citenamefont {Ledwith}, \citenamefont {Watanabe}, \citenamefont {Taniguchi}, \citenamefont {Khalaf}, \citenamefont {Vishwanath},\ and\ \citenamefont {Efetov}}]{Shen2023}%
  \BibitemOpen
  \bibfield  {author} {\bibinfo {author} {\bibfnamefont {C.}~\bibnamefont {Shen}}, \bibinfo {author} {\bibfnamefont {P.~J.}\ \bibnamefont {Ledwith}}, \bibinfo {author} {\bibfnamefont {K.}~\bibnamefont {Watanabe}}, \bibinfo {author} {\bibfnamefont {T.}~\bibnamefont {Taniguchi}}, \bibinfo {author} {\bibfnamefont {E.}~\bibnamefont {Khalaf}}, \bibinfo {author} {\bibfnamefont {A.}~\bibnamefont {Vishwanath}},\ and\ \bibinfo {author} {\bibfnamefont {D.~K.}\ \bibnamefont {Efetov}},\ }\href {https://doi.org/10.1038/s41563-022-01428-6} {\bibfield  {journal} {\bibinfo  {journal} {Nature Materials}\ }\textbf {\bibinfo {volume} {22}},\ \bibinfo {pages} {316} (\bibinfo {year} {2023})}\BibitemShut {NoStop}%
\bibitem [{\citenamefont {Christos}\ \emph {et~al.}(2022)\citenamefont {Christos}, \citenamefont {Sachdev},\ and\ \citenamefont {Scheurer}}]{PhysRevX.12.021018}%
  \BibitemOpen
  \bibfield  {author} {\bibinfo {author} {\bibfnamefont {M.}~\bibnamefont {Christos}}, \bibinfo {author} {\bibfnamefont {S.}~\bibnamefont {Sachdev}},\ and\ \bibinfo {author} {\bibfnamefont {M.~S.}\ \bibnamefont {Scheurer}},\ }\href {https://doi.org/10.1103/PhysRevX.12.021018} {\bibfield  {journal} {\bibinfo  {journal} {Phys. Rev. X}\ }\textbf {\bibinfo {volume} {12}},\ \bibinfo {pages} {021018} (\bibinfo {year} {2022})}\BibitemShut {NoStop}%
\bibitem [{\citenamefont {Han}\ \emph {et~al.}(2024)\citenamefont {Han}, \citenamefont {Herzog-Arbeitman}, \citenamefont {Bernevig},\ and\ \citenamefont {Kivelson}}]{Geo_flat_band_Han2024}%
  \BibitemOpen
  \bibfield  {author} {\bibinfo {author} {\bibfnamefont {Z.}~\bibnamefont {Han}}, \bibinfo {author} {\bibfnamefont {J.}~\bibnamefont {Herzog-Arbeitman}}, \bibinfo {author} {\bibfnamefont {B.~A.}\ \bibnamefont {Bernevig}},\ and\ \bibinfo {author} {\bibfnamefont {S.~A.}\ \bibnamefont {Kivelson}},\ }\bibfield  {journal} {\bibinfo  {journal} {Physical Review X}\ }\textbf {\bibinfo {volume} {14}},\ \href {https://doi.org/10.1103/physrevx.14.041004} {10.1103/physrevx.14.041004} (\bibinfo {year} {2024})\BibitemShut {NoStop}%
\bibitem [{\citenamefont {Gani}\ \emph {et~al.}(2019)\citenamefont {Gani}, \citenamefont {Steinberg},\ and\ \citenamefont {Rossi}}]{SC_hetero_Gani2019}%
  \BibitemOpen
  \bibfield  {author} {\bibinfo {author} {\bibfnamefont {Y.~S.}\ \bibnamefont {Gani}}, \bibinfo {author} {\bibfnamefont {H.}~\bibnamefont {Steinberg}},\ and\ \bibinfo {author} {\bibfnamefont {E.}~\bibnamefont {Rossi}},\ }\bibfield  {journal} {\bibinfo  {journal} {Physical Review B}\ }\textbf {\bibinfo {volume} {99}},\ \href {https://doi.org/10.1103/physrevb.99.235404} {10.1103/physrevb.99.235404} (\bibinfo {year} {2019})\BibitemShut {NoStop}%
\bibitem [{\citenamefont {Xie}\ and\ \citenamefont {Law}(2023)}]{PhysRevLett.131.016001}%
  \BibitemOpen
  \bibfield  {author} {\bibinfo {author} {\bibfnamefont {Y.-M.}\ \bibnamefont {Xie}}\ and\ \bibinfo {author} {\bibfnamefont {K.~T.}\ \bibnamefont {Law}},\ }\href {https://doi.org/10.1103/PhysRevLett.131.016001} {\bibfield  {journal} {\bibinfo  {journal} {Phys. Rev. Lett.}\ }\textbf {\bibinfo {volume} {131}},\ \bibinfo {pages} {016001} (\bibinfo {year} {2023})}\BibitemShut {NoStop}%
\bibitem [{\citenamefont {Qin}\ and\ \citenamefont {MacDonald}(2021)}]{PhysRevLett.127.097001}%
  \BibitemOpen
  \bibfield  {author} {\bibinfo {author} {\bibfnamefont {W.}~\bibnamefont {Qin}}\ and\ \bibinfo {author} {\bibfnamefont {A.~H.}\ \bibnamefont {MacDonald}},\ }\href {https://doi.org/10.1103/PhysRevLett.127.097001} {\bibfield  {journal} {\bibinfo  {journal} {Phys. Rev. Lett.}\ }\textbf {\bibinfo {volume} {127}},\ \bibinfo {pages} {097001} (\bibinfo {year} {2021})}\BibitemShut {NoStop}%
\end{thebibliography}
\end{document}